\documentclass[journal,onecolumn,draftcls]{IEEEtran}
\usepackage{amsmath,amsfonts,amssymb,amsbsy, amsthm,ae,aecompl}
\usepackage{algorithm,algpseudocode}
\usepackage[utf8]{inputenc}
\usepackage[english]{babel}
\usepackage{bm}
\usepackage{color}
\usepackage{epsfig}
\usepackage{enumerate}
\usepackage{float}
\usepackage{graphicx}
\usepackage{graphics}
\usepackage{tcolorbox}
\usepackage{fancyhdr}
\usepackage[T1]{fontenc}
\usepackage[acronym,toc,shortcuts]{glossaries}
\usepackage{epstopdf}
\usepackage{hyperref}
\usepackage{lastpage}
\usepackage{ulem}
\usepackage{listings}
\usepackage{lipsum}
\usepackage{dsfont}
\usepackage{amsmath,amsfonts}
\usepackage{verbatim}
\usepackage[all]{xy}
\newtheorem{theorem}{Theorem}

\usepackage{bbm}
\usepackage[caption=false]{subfig}
\DeclareMathOperator*{\argmax}{arg\,max}

\begin{document}
\title{Ultra-Reliable and Low-Latency Wireless Communication:  Tail, Risk and Scale}
%\author{Mehdi~Bennis, Senior Member  IEEE,  Mérouane~Debbah,  Fellow IEEE, and others}
\author{Mehdi~Bennis, Senior Member  IEEE,  Mérouane~Debbah,  Fellow IEEE, and H. Vincent Poor, Fellow IEEE
\thanks{M. Bennis is with the Centre for wireless communications, University of Oulu, 4500 Oulu, Finland, (e-mail: mehdi.bennis@oulu.fi). M. Debbah is with the Mathematical and Algorithmic Sciences Lab, Huawei France R$\&$D, Paris, France, (e-mail: merouane.debbah@huawei.com). H. V. Poor is with the Department of Electrical Engineering, Princeton University, Princeton, NJ 08544 USA (e-mail: poor@princeton.edu).}

}
        % <-this % stops a space

\maketitle
%\vspace{-100ex}
%\IEEEpeerreviewmaketitle

%%%%%%%%%%%%%%%%%%%%%%%%%%%%%%%%%%%%%%%%%%%%%%%%%%%%%%%%%%%%%%%%%%%
%%%%%%%%%%%%%%%%%%%%%%%%%%%%%%%%%% ABSTRACT
\begin{abstract}
%We are on the cusp of a disruptive world with billions of connected devices communicate, with new experiences such as connected cities, autonomous cars, virtual and augmented reality. 5G will bring about unprecedented  requirements for speed, latency, reliability, energy and scale.   However, to make this a reality incremental changes to the way networks have been optimized are obsolete calling for a truly new paradigm change.

Ensuring ultra-reliable  and low-latency communication (URLLC) for  5G wireless networks  and beyond is of capital importance and is currently receiving tremendous attention in academia and industry.  At its core, URLLC mandates a departure from expected utility-based network design approaches, in which relying on average quantities (e.g., average throughput, average delay and average response time) is no longer an option but a necessity. Instead, a principled and scalable framework which takes into account delay, reliability, packet size, network architecture, and topology (across access, edge, and core) and decision-making under uncertainty is sorely lacking.  The overarching goal of this article is a first step to fill this void. Towards this vision,  after  providing  definitions of latency and reliability,  we closely examine various enablers of URLLC and their inherent tradeoffs. Subsequently, we focus our attention on a plethora of techniques and methodologies pertaining to the requirements of ultra-reliable and low-latency communication, as well as their  applications through selected use cases. These results provide crisp insights for the design of low-latency and high-reliable wireless networks.
\end{abstract}


\section{Introduction}

The phenomenal growth of data traffic spurred by the internet-of-things (IoT) applications ranging from machine-type communications (MTC) to mission-critical communications (autonomous driving, drones and augmented/virtual reality) are posing unprecedented challenges in terms of capacity, latency, reliability, and scalability  \cite{BastugBMD16} \cite{Popovski14},\cite{7412759}, \cite{IIOT}.  This is further exacerbated by: i) a growing network size and increasing interactions between nodes; ii) a high level of uncertainty due to random changes in the topology; and iii) a heterogeneity across applications, networks and devices. The stringent requirements of these new applications warrant a paradigm shift  from reactive and    centralized networks towards massive, low-latency, ultra-reliable and proactive 5G networks. Up until now,  human-centric communication networks have been engineered with a focus on improving network capacity with little attention to latency or reliability, while assuming few users.

Achieving ultra-reliable and low-latency communication (URLLC) represents one of the major challenges facing 5G networks.  URLLC introduces a plethora of challenges in terms of system design. While enhanced mobile broadband (eMBB) aims at high spectral efficiency, it can also rely on hybrid automatic repeat request (HARQ) retransmissions to achieve high reliability. This is, however, not the case for URLLC due to the hard latency constraints. Moreover, while ensuring URLLC at a link level in controlled environments is relatively easy, doing it at a network level and over a  wide area and in remote scenarios (e.g., remote surgery \cite{5gppp,BJU4475}) is notoriously difficult. This is due to the fact that for local area use cases latency is mainly due to the wireless media access, whereas wide area scenarios suffer from latency due to intermediate nodes/paths, fronthaul/backhaul and the core/cloud. Moreover, the typical block error rate (BLER) of 4G systems is $10^{-2}$ which can be achieved by channel coding (e.g., Turbo code) and re-transmission mechanisms (e.g., via HARQ). By contrast, the performance requirements of URLLC are more stringent with a target BLER of  $[10^{-9}-10^{-5}]$ depending on the use case \cite{3gppreq}. From a physical-layer perspective, the URLLC design is challenging as it ought to satisfy two conflicting requirements: low latency and ultra-high reliability \cite{7063628}. On the one hand,  minimizing latency mandates the use of short packets which in turns causes a severe degradation in channel coding gain \cite{DurisiKP16,7529226}. On the other hand, ensuring reliability requires more resources (e.g., parity, redundancy, and re-transmissions) albeit increasing latency (notably for time-domain redundancy). Furthermore,  URLLC warrants a system design tailored to the unique requirements of different verticals for which the outage capacity is of interest (as opposed to the Ergodic capacity considered in 4G \cite{7980747}). This ranges from users (including cell edge users) connected to the radio access network which must  receive equal grade of service \cite{7980747}, to vehicles reliably transmitting their safety messages \cite{Ashraf} and industrial plants whereby sensors, actuators and controllers communicate within very short cycles \cite{LuvisottoPD17}.

\begin{figure} [h!]
 \includegraphics[width=\columnwidth]{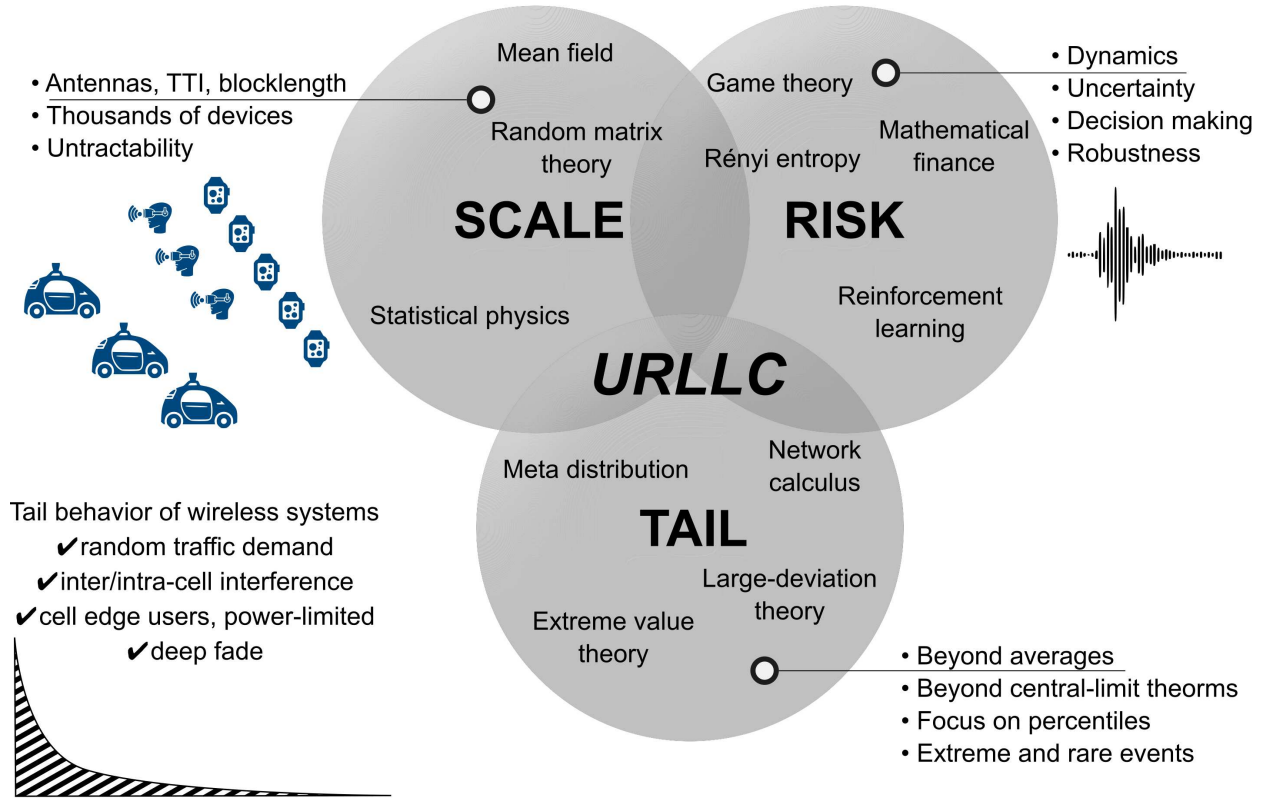}
\caption{Anatomy  of the  URLLC building blocks, composed of tail, scale and risk alongside their unique characteristics. In addition, mathematical tools tailored to the  requirements of every block are highlighted therein.}
 \end{figure}

If successful, URLLC will unleash a plethora of novel applications and digitize a multitude of verticals. For instance, the targeted $1$ ms latency time (and even lower) is crucial in the use of haptic feedback and real-time sensors to allow doctors to examine patients' bodies from a remote operating room \cite{BJU4475}. Similarly, the construction industry can operate heavy machinery remotely and minimize other potential  hazards \cite{7403840}.  For sports fans, instead of watching   NBA games on television, using a virtual reality (VR) headset allows to  have a  $360$-degree courtside view, feeling the intensity of the crowd from the comfort of the home \cite{BastugBMD16}. The end-to-end latency for XR (augmented, virtual and immersive reality) represents a serious challenge that ought to be tackled.  Likewise,  ultra-high reliability in terms of successful packet rate delivery, which may be as high as $1-10^{-5}$  (or  even $ 1-10^{-9}$), will help automate factories, spearhead remote monitoring and control \cite{IIOT}. Undoubtedly, these technological advances will only be possible with a scalable,  ultra-reliable and low-latency network.

\noindent In essence, as shown in Figure 1,   URLLC can be broken down into three major building blocks, namely: (i) risk, (ii) tail, and  (iii) scale.
 \begin{itemize}
  \item \textbf{Risk}: risk is naturally encountered when dealing with decision making under uncertainty, when channels are time-varying and in the presence of network dynamics. Here, decentralized or semi-centralized algorithms providing performance guarantees and robustness are at stake, and notably game theory and reinforcement learning. Furthermore, Renyi entropy is an information-theoretic criterion that precisely quantifies  uncertainty embedded in a distribution, accounting for all moments, encompassing Shannon entropy as a special case \cite{renyi1961}.
\item \textbf{Tail}:  the notion of tail behavior in  wireless systems is inherently related to the tail of random traffic demand, tail of latency distribution, intra/inter-cell interference, and users that are at the cell edge, power-limited or in deep fade. Therefore, a principled framework and mathematical tools that characterize these tails focusing on percentiles and extreme events are needed. In this regard, extreme value theory \cite{Coles01}, mathematical finance and network calculus are  important methodologies.
    \item \textbf{Scale}: this is motivated by the sheer amount of devices, antennas, sensors and other nodes which pose serious challenges in terms of resource allocation and network design.  Scale is highly relevant for  mission-critical machine type communication use cases (e.g.,  industrial process automation) which are based on a large number of sensors, actuators and controllers  requiring communication with very high reliability and low end-to-end latency \cite{IIOT}, \cite{7497764}\footnote{Table $1$ shows the scaling, reliability, latency and range for a wireless factory use case \cite{7497764}.}. In contrast to cumbersome and time-consuming    Monte-Carlo simulations, mathematical tools focused on large system analysis which provide a tractable formulation  and  insights are needed. For this purpose, mean field (and mean field game) theory \cite{book:gueant11}, statistical physics and random matrix theory \cite{Couillet} are important tools.
\end{itemize}

%\begin{figure}
 %\includegraphics[width=\textwidth]{Picture2.eps}
%\caption{Various denominations of Ultra Reliability and Low latency: URC, LLC and URLLC.}
 %\end{figure}

%In what follows, we have identified a (non-exhaustive) set of tools and methodologies which serve this purpose:
The article is structured as follows: to guide  the  readers and set the stage for the technical  part,   definitions of latency and reliability are presented in Section II. Section III provides a state-of-the-art summary of the most recent and relevant works.  Section IV delves into the details of some of the key enablers of URLLC, and section V examines several  tradeoffs cognizant of the URLLC characteristics. Next, Section VI presents an overview of various tools and techniques tailored to the unique features of URLLC (risk, scale and tail). Finally,  we illustrate through selected use cases the usefulness of some of the methodologies in Section VII followed by concluding remarks.

\section{Definitions}
\subsection{Latency}
\begin{itemize}
  \item \textbf{End-to-end (E2E) latency \cite{3gpplatency}}: E2E latency  includes the over-the-air transmission delay, queuing delay, processing/computing delay and retransmissions, when needed. Ensuring a  round-trip  latency of $1$ ms and owing to the speed of light constraints ($300$ km/ms), the maximum distance at which a receiver can be located is approximately $150$ km.
  \item	\textbf{User plane latency (3GPP) \cite{3gpp}}: defined as the one-way time it takes to successfully deliver an application layer packet/message from the radio protocol layer ingress point to the radio protocol ingress point of the radio interface, in either uplink or downlink in the network for a given service in unloaded conditions (assuming the user equipment (UE) is in active state). The minimum requirements for user plane latency are  4 ms for eMBB and  1 ms for URLLC assuming  a single user.
\item 	\textbf{Control plane latency (3GPP)  \cite{3gpp}}: defined as the transition time from a most “battery efficient” state (e.g., idle state) to the start of continuous data transfer (e.g. active state). The minimum requirement for control plane latency is 20 ms.
\end{itemize}

\subsection{Reliability}
  In general, reliability is defined as the probability that a data of size $D$ is successfully transferred within a time period $T$. That is, reliability stipulates that  packets are successfully delivered and the latency bound is satisfied. However, other definitions can be encountered:

\begin{itemize}
\item	\textbf{Reliability (3GPP)  \cite{3gpp}}:  capability of transmitting a given amount of traffic within a predetermined time duration with high success probability. The minimum requirement for  reliability is $1-10^{-5}$ success probability of transmitting a layer 2 protocol data unit of 32 bytes within 1 ms.
%\item	\textbf{Reliability (ITU) \cite{ITU}}:  probability to guarantee a required function/performance under stated conditions for a given time interval. The specific reliability requirements differ for various types of services and applications.
  \item \textbf{Reliability per node}:  defined as the transmission error probability, queuing delay violation probability and proactive packet dropping probability.
\item \textbf{Control channel reliability}: defined as the probability of successfully decoding the scheduling grant or  other metadata \cite{JiPYKLS17}.
\item \textbf{Availability}:  defined as the probability that a given service is available (i.e., coverage). For instance,   $99.99\%$ availability means that one user among 10000 does not receive  proper coverage \cite{7063630}.
\end{itemize}

\noindent We underscore the fact that  URLLC service requirements  are end-to-end, whereas the 3GPP and ITU requirements focus on the \emph{one-way radio latency}  over the 5G radio network \cite{3gpp}.

\section{State-of-the-art and gist of recent work}

Ensuring low-latency and ultra-reliable communication for future wireless networks is of capital importance. To date, no   work has been done on combining latency and reliability into a theoretical framework, although the groundwork has been laid by Polyanskiy’s development of bounds on block error rates for finite blocklength codes \cite{5452208,7156144}. These works are essentially information-theoretic and overlook queuing effects and other networking issues.  Besides that, no wireless communication systems have been proposed with latency constraints on the order of milliseconds with  hundreds to thousands of nodes and with system reliability requirements of $1-10^{-6}$ to $1-10^{-9}$.

\subsection{Latency}

At the physical layer level, low-latency communication has been studied  in terms of  throughput-delay tradeoffs \cite{1354518}. Other theoretical investigations include delay-limited link capacity \cite{737514} and the use of network effective capacity \cite{515976}.  While interesting these works focus on minimizing the average latency instead of the worst-case latency. At the network level, the literature on queue-based resource allocation is rich in which tools from Lyapunov optimization,  based on  myopic queue-length based optimization   are state-of-the-art \cite{Neely13}. However, while stability is an important aspect of queuing networks, fine-grained metrics like the delay distribution and probabilistic bounds (i.e., tails) cannot be addressed. Indeed a long-standing challenge is to  understand the non-asymptotic trade-offs between delay, throughput and reliability in wireless networks including both coding  and queuing delays. Towards this vision, the recent works of Al-Zubaidy  et al. \cite{mellin} constitute a very good starting point, leveraging the framework of stochastic network calculus. Other interesting  works geared towards latency reduction at the network level include   caching at the network edge (base stations and user equipment) \cite{BastugBZKKED15,edge01, BastugBD15, PantisanoBSD14, edge01,7435255,Femtocaching},  the use of shorter transmission time interval (TTI) \cite{7980747},  grant-free based non orthogonal multiple access \cite{7600901,7063628} and mobile edge computing \cite{GC17_EVT, SabellaVKRG16}, to mention a few.

\subsection{Reliability}
Reliable communication has been a fundamental problem in information theory since Shannon's seminal paper showing that it is possible to communicate with vanishing probability of error at non-zero rates \cite{Shannon01}. Several decades after saw the advent of many error control coding  schemes for point-to-point communication (Turbo, LDPC and Polar codes) \cite{DurisiKP16}. In wireless fading channels, diversity schemes were developed to deal with the deep fades stemming from multipath effects. For coding delays, error exponents (reliability functions) characterize the exponential rates at which error probabilities decay as coding block-lengths become large \cite{NakibogluG08}. However, this approach does not capture sub-exponential terms needed to characterize the low delay performance (i.e. the tails). Recent works on finite-block length analysis and channel dispersion \cite{5452208,7156144,DurisiKP16} help in this regard but do not address multiuser wireless networks nor  interference-limited settings. At the network level, reliability has  been studied to complement the techniques used  at the physical layer, including automatic repeat request (ARQ) and its hybrid version (HARQ) at  the medium access level. In these works, reliability is usually increased at the cost of latency due to the use of longer blocklengths or through the use of retransmissions.

Recently, packet duplication was shown to achieve high reliability in \cite{Popovski14,DurisiKP16},  high availability using multi-connectivity was studied in  \cite{7063630}  in an interference-free scenario and stochastic network calculus was   applied in a single user and multiple input single output (MISO) setting in \cite{ArnauK17aa}. In terms of 5G verticals, challenges of ultra-reliable  vehicular communication were looked at in \cite{Ashraf, abs-1712-00537}, whereas   mobile edge computing with URLLC guarantees  was studied in \cite{ElBambyBS17}.  Network slicing for enabling ultra-reliable industrial automation was examined in \cite{IIOT}, and  a maximum average rate was derived in \cite{ParkP17} guaranteeing a signal-to-interference ratio. Finally, a  recent (high-level) URLLC survey article can be found in \cite{abs-1708-07862} highlighting the building principles of URLLC.

\noindent \textbf{Summary.}  Most  of the current state-of-the-art  has made a significant contribution towards understanding the ergodic capacity and the average queuing performance of wireless networks focusing on large blocklength. However, these  works fall short of providing  insights for reliability and latency issues and  understanding  their non-asymptotic tradeoffs. In addition, current radio access networks  are designed with the aim of maximizing throughput while considering a few active users. Progress has been made in URLLC in terms of short-packet transmission \cite{DurisiKP16, 5452208} and other concepts such as shorter transmission time interval, network slicing, caching, multi-connectivity, and so forth. However, a principled framework laying down the fundamentals of URLLC at the network level, that is scalable and features a system design centered on tails is  lacking.

\section{Key Enablers for URLLC}
In this section, key enablers for low-latency and high-reliability communication are examined. An overview of some of these enablers is highlighted in Figure 2, while  Table~\ref{tab:a} provides a comparison between 4G and 5G.
\subsection{Low-Latency}
A latency  breakdown yields deterministic and random components that are either fixed or scale with the number of nodes.  While the deterministic component defines the minimum latency,  the random components impact the latency distribution and more specifically its  tails. Deterministic latency components consist of the time to transmit information and overhead (i.e. parity bits, reference signals and control data),  and waiting times between transmissions. The random components include the time to retransmit information and overhead when necessary, queuing delays, random backoff times,  and other processing/computing delays. In what follows, various enablers for low-latency communication are examined:

\begin{itemize}
\item	\textbf{Short transmission time interval (TTI), short frame structure and hybrid automatic repeat request(HARQ)}:  reducing the TTI duration (e.g., from $1$ ms in LTE to $0.125$ ms as in 5G new radio) using fewer OFDM\footnote{$2$ OFDM symbols = $71.43$ microseconds with  a spacing of $30$ KHz.} symbols per TTI and  shortening OFDM symbols via wider subcarrier spacing as well as lowering HARQ roundtrip time (RTT) reduce latency. This is because less time is needed to have enough HARQ retransmissions to meet a reliability target and tolerate more queuing delay before the deadline (owing to the HARQ retransmissions constraints). Furthermore, reducing the OFDM symbol duration increases subcarrier spacing and hence fewer resource blocks are available in the frequency domain causing more queuing effect.  On the flipside, shorter TTI duration introduces more control overhead thereby reducing capacity (lower availability of resources for other URLLC data transmissions). This shortcoming can be alleviated using grant-free transmission in the uplink. In the downlink, longer TTIs are needed at high offered loads  to cope with non-negligible queuing delays \cite{7980747}.
\item	\textbf{eMBB/URLLC multiplexing}: Although a static/semi-static resource partitioning between eMBB and URLLC transmissions may be preferable in terms of latency/reliability viewpoint, it is inefficient in terms of system resource utilization, calling for a dynamic multiplexing solution \cite{7980747}.	Achieving high system reliability for URLLC requires more frequency-domain resources to be allocated to an uplink (UL) transmission instead of boosting power on narrowband resources. This means wideband resources are needed for URLLC UL transmission to achieve high reliability with low latency. In addition, intelligent scheduling techniques to preempt other scheduled traffic are needed when  a low-latency  packet arrives in the middle of the frame (i.e.,  puncturing the current eMBB transmission). At the same time, the eMBB traffic should be minimally impacted  when maximizing the URLLC outage capacity. A recent work towards this direction can be found in \cite{shakkotai} as well as a recent application in the context of virtual reality \cite{abs-1805-00142}.
\item	\textbf{Edge caching,  computing and slicing}: pushing caching and computing resources to the network edge has been shown to significantly reduce latency \cite{BastugBZKKED16,edge}. This trend will continue unabated with the advent of resource-intensive applications (e.g., augmented and virtual reality) and other mission-critical applications (e.g., autonomous driving). In parallel to that, network slicing will also play a pivotal role in allocating dedicated   caching, bandwidth and computing resources (slices).
    \item	\textbf{On-device machine learning and artificial intelligence (AI) at the network edge}:     Machine learning (ML) lies at the foundation of proactive and low-latency networks. Traditional ML is based on the precept of a single node (in a centralized location) with access to the global dataset and a massive amount of storage and computing, sifting through this data for classification and inference. Nevertheless, this approach is clearly inadequate for latency-sensitive and high-reliability applications, sparking a huge interest in distributed ML (e.g, deep learning coined by Lecun et. al. \cite{LeCunBH15}). This mandates a novel scalable and distributed machine learning framework, in which the training data describing the problem is stored in a distributed fashion across a number of interconnected nodes and the optimization problem is solved collectively. This constitutes the next frontier for ML, also referred to as \emph{AI-on-edge} or \emph{on-device ML} \cite{KonecnyMRR16}. Our preliminary work in this direction can be found in \cite{FLBENNIS}.

\item	\textbf{Grant-free versus grant-based access}:  This relates to dynamic uplink scheduling  or contention-based access for sporadic/bursty traffic versus persistent scheduling for periodic traffic. Fast uplink access is advocated for devices on a priori basis at the expense of lower capacity (due to resource pre-allocation). For semi-persistent scheduling unused resources can be reallocated to the eMBB traffic. In another solution referred to as  group-based semi-persistent scheduling, contention based access is carried out within a group of users with similar characteristics, thereby  minimizing collisions within the group \cite{7925877}. In this case the base station  controls the load and dynamically adjusts the size of the resource pool. For retransmissions, the BS could also proactively schedule a retransmission opportunity shared by a group of UEs with similar traffic for better resource utilization. On the other hand, grant-free access shortens the procedure for uplink resource assignment, whereby the  reservation  phase is skipped.
\item	\textbf{Non-orthogonal multiple access (NOMA)}: NOMA (and its variants)  reduces latency by supporting far more users than  conventional orthogonal-based approaches  leveraging power or code domain multiplexing in the uplink then using   successive interference cancellation (SIC), or more advanced receiver schemes (e.g., message passing or Turbo reception). However, issues related to imperfect channel state information (CSI), user ordering, processing delay due to multiplexing and other dynamics which impact latency (and reliability) are  not well understood.
\item	\textbf{Low/Medium-earth orbit (LEO/MEO) satellites and unmanned aerial vehicles (UAVs)}: Generally,  connecting terrestrial base  stations  to  the  core  network  necessitates wired  backhauling\footnote{In addition, backhaul provisioning can be done by leveraging massive MIMO, whereby instead of deploying more base stations, some antenna elements can be used for wireless backhauling \cite{VuBSDL17,SamarakoonBSL13}.}. However,  wired  connections  are expensive  and sometimes infeasible  due to geographical constraints such as remote areas. In this  case,  UAVs   can enable a   reliable  and  low-latency wireless backhaul connectivity for ground networks \cite{7412759, ChallitaS17,UAVMF}. Furthermore, for long-range applications or in rural areas, LEO satellites are the only way to reduce backhaul latency in which a hybrid architecture composed of balloons, LEOs/MEOs and other stratospheric vehicles are low-latency communication enablers \cite{OuimetCM15}.
\item \textbf{Joint flexible resource allocation for uplink/downlink}: for time division duplex (TDD) systems, a joint uplink/downlink allocation and the interplay of time slot length versus the switching cost (or turn around) is needed. This topic has been studied in the context of LTE-A. Here, for frequency division duplex (FDD)  both LTE evolution and the new radio (NR) are investigated, whereas for TDD only NR is investigated since LTE TDD is not considered for URLLC enhancements.
\end{itemize}

\begin{figure} [h!]
  \includegraphics[width=\textwidth]{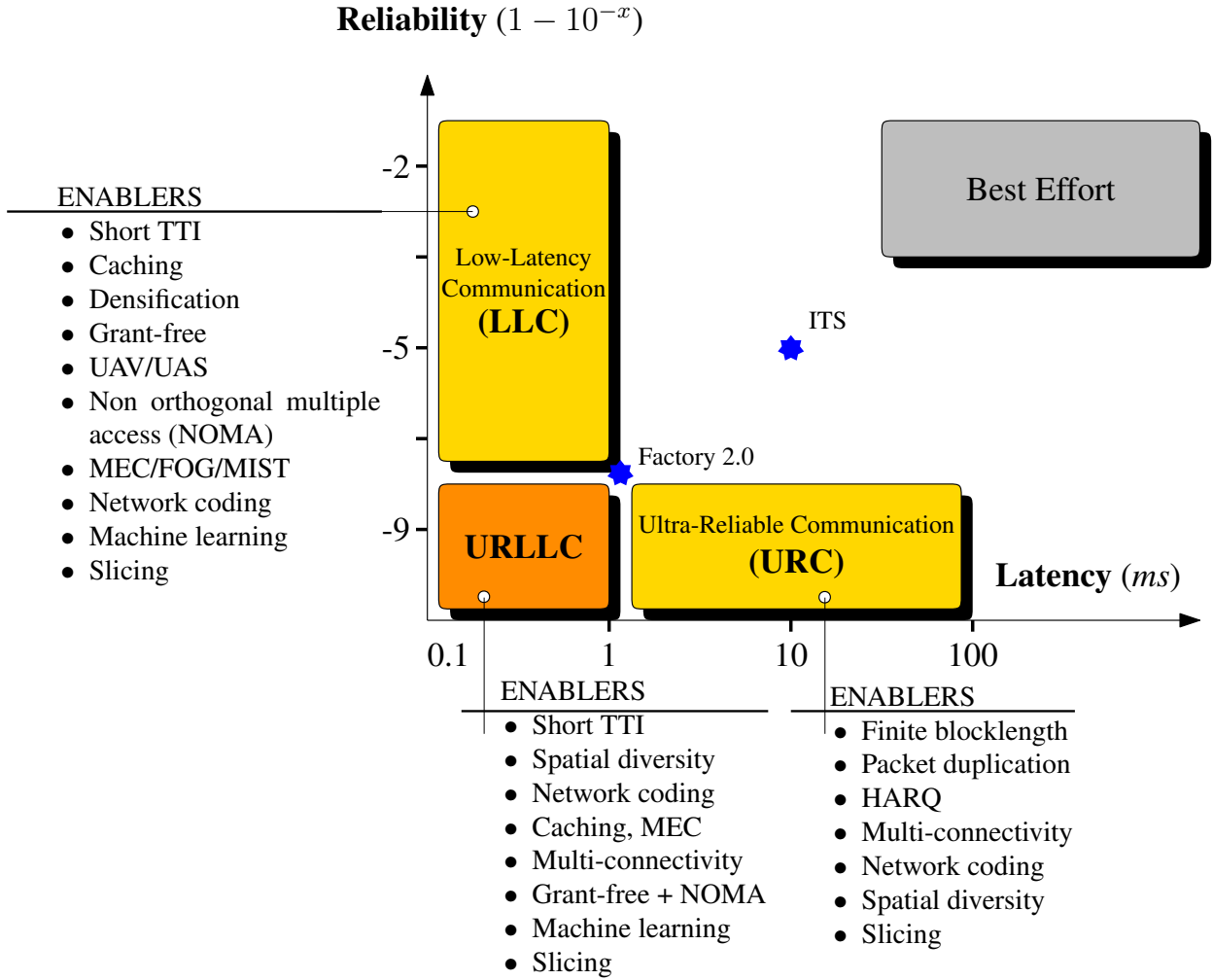}
\caption{A breakdown of the key enablers for low latency and high reliability.}
 \end{figure}

\subsection{Reliability}
The main factors affecting  reliability stem from: i) collisions with other users due to uncoordinated channel access; ii) coexistence with other systems in the same frequency bands; iii) interference from users in adjacent channels; iv) Doppler shifts from moving devices, v) difficulty of synchronization,  outdated channel state information, time-varying channel effects or delayed packet reception. Reliability at the physical layer level (typically expressed in  block error rate) depends on  factors such as the channel, constellation, error detection codes, modulation technique, diversity, retransmission mechanisms, etc. A variety of techniques to increase reliability include  using  low-rate codes to have enough redundancy in poor channel conditions, retransmissions for error correction, and  ARQ at the transport layer.  Crucially, diversity and beamforming provide multiple independent paths from the transmitter to the receiver and to boost the received signal-to-noise ratio (SNR). Frequency diversity occurs when information is transmitted over a frequency-selective channel whereas time diversity occurs when a forward error correction (FEC) codeword is spread out over many coherence times so that it sees many different channels (for e.g., using HARQ). Multi-user diversity arises when a transmission is relayed by different users from the source to the destination. In what follows, various enablers for reliability are discussed.

\begin{itemize}
\item	\textbf{Multi-connectivity  and harnessing time/frequency/RATs diversity}:  while diversity is  a must, time diversity is not a viable solution when the tolerable latency is shorter than the channel coherence time  or when the reliability requirements are very stringent. On the other hand frequency diversity may not scale with the number of users/devices, making spatial diversity the only solution. In this regard, multi-connectivity is paramount in ensuring high-reliable communication. Nevertheless several fundamental questions arise, such as  what is the optimal number of links  needed to ensure a given reliability target? issues of correlated links versus independent links; how to deal with synchronization, non-reciprocity issues, and other imperfections?
\item \textbf{Multicast}: when receivers are interested in the same information  (e.g., mission-critical traffic safety or a common field-of-view in virtual reality), multicast is more reliable than using unicast. However reliability can be sensitive to the  coverage range, what modulation and coding scheme (MCS)  is used within the multicast group and who determines the MCS? Moreover, the usefulness of multicast will  depend  on whether transmission is long-range or short-range as the performance is inherently limited by cell edge users.
\item	\textbf{Data replication (contents and computations)}: needed when coordination among nodes is not possible, when a low-rate backhaul is needed for coordination or due to lack of channel state information. This comes at the expense of lower capacity. One solution could be to replicate the same data until receiving an acknowledgement in the case of HARQ.
\item	\textbf{HARQ + short frame structure, short TTI}: improve outage capacity via sufficient retransmissions to achieve high reliability.  Here an optimal MCS level selection with the constraint of required reliability and latency (not necessarily optimized for spectral efficiency) is an open research problem.
    \item	\textbf{Control channel design}:   Unlike LTE where the focus was primarily on protecting data but not the control channel,   ensuring high reliability for the control channel is a must. This can be done by sending the delay budget information from the user to the base station (BS) in the control channel, such that on the downlink the BS can select the optimal modulation coding scheme based on both channel quality indicator (CQI)\footnote{Adaptive channel quality indicator (CQI) reporting is key in URLLC whereby more resources can be allocated to enhance the reliability or changing the encoding rate of the CQI.} report and remaining latency budget. In addition, replicating the same data until receiving an acknowledgement for HARQ can be envisaged at the cost of wasting resources.
%\item	\textbf{Manufacturing diversity via network coding and relaying}: In cases when time diversity cannot be relied upon or in the presence of extreme fading events, manufacturing diversity and robustness is a must.   Hence exploiting multi-user diversity and network coding using simultaneous relaying to enable a two way reliable communication  without relying on time and frequency diversity is important.
    \item	\textbf{Manufacturing diversity via network coding and relaying}: when time diversity cannot be relied upon (due to extreme latency and reliability constraints) or in the presence of extreme fading events, manufacturing diversity and robustness are key for ensuring URLLC.   For instance, in industrial automation  exploiting multi-user diversity and network coding using simultaneous relaying to enable a two-way reliable communication  without relying on time and frequency diversity is important due to the cycle time constraints, channel and network dynamics \cite{IIOT}.
        %Furthermore, network densification in terms of antennas, base stations or relays not only reduces latency by shrinking the transmission range, but also increases capacity and reliability.  Here, depending on the use case of interest, this may come at the expense of backhaul provisioning (for e.g., when deploying more base stations), in which case adding more antenna elements per BS and leveraging massive MIMO is preferable  than deploying more base stations \cite{ArnauK17aa}.}

\item	\textbf{Network slicing}: it refers to the process of slicing a physical network into logical sub-networks  optimized  for specific applications, in which the goal is to allocate dedicated resources for  verticals of interest, such as vehicle-to-vehicle communication and industrial automation \cite{7589996}.  Constructing  network  slices  with  stringent  reliability  guarantees for these mission-critical applications is  a daunting task  due  to  the  difficulties  in  modeling  and  predicting  queuing delays with very high accuracy.  For instance, in the industrial automation scenario, owing to its diverse set of application requirements, \emph{one} slice could be allocated for a high-definition  video transmission between a person remotely monitoring a process (or robot), and \emph{another} slice to provide an ultra-reliable transmission between sensors, controllers and actuators \cite{IIOT,LuvisottoPD17}.
\item	\textbf{Proactive packet drop}: when channel is in a deep fade,  packets that cannot be transmitted even with the maximal transmit power can be discarded proactively at the transmitter. Similarly, packet drop can arise at the receiver when the maximum number of re-transmissions is reached. This is different than the enhanced mobile broadband scenarios based on infinite queue buffers. In this case either spatial diversity should be used or resources need to be increased.
    \item	\textbf{Space-time block codes}: Orthogonal space-time block coding has been a very successful transmit diversity technique because it  achieves full diversity without CSI at the transmitter and need for joint decoding of multiple symbols. Typically, it  is characterized by the number of independent symbols $N_s$ transmitted over $T$ time slots; the code rate is $R_c = N_s/T$. In the presence of channel imperfection, orthogonal space-time block coding can outperform other diversity-seeking approaches such as the maximum ratio transmission.
\end{itemize}

\begin{table}[]
\centering
\caption{4G vs. 5G (in a nutshell)}
\label{tab:a}
\begin{tabular}{|l|l|l|}
\hline
              & \textbf{4G}          & \textbf{5G}      \\ \hline
\textbf{Metadata}        & important & crucial  \\ \hline
\textbf{Packet size}     & Long (MBB)            &   Short (URLLC)- Long (eMBB)      \\ \hline
\textbf{Design}          &    Throughput-centric, Average delay good enough
         &    Latency and reliability centric / tails MATTER      \\ \hline
\textbf{Reliability}     &  95$\%$ or less
           &        $1-10^{-x}$       $x=[3,4,5,6,8,9]$  \\ \hline
\textbf{Rate}            &   Shannonian (long packets)          &       Rate loss due to short packets    \\ \hline
\textbf{Delay violation} &    Exponential decay using effective bandwidth         &       Faster decay than exponential     \\ \hline
\textbf{Latency}         &  15 ms RTT based on 1 ms subframe           &       1 ms and less, shorter TTI, HARQ RTT    \\ \hline
\textbf{Queue size }     &  unbounded           &     bounded     \\ \hline
\textbf{Frequency bands} &    sub-6GHz        &       Above- and sub-6GHz (URLLC at sub-6GHz)     \\ \hline
\textbf{Scale}           &   A few users/devices          & billion devices  \\ \hline

\end{tabular}
\end{table}

\section{Fundamental Trade-offs in URLLC}

%One of the biggest distinguishing factors of mission-critical applications are the requirements with respect to the delay. For instance, in factory automation  closed-loop control systems, where sensors, controllers, and actuators must exchange information with cycle times (i.e. delays) of few ms and reliability levels of $10^{-9}$. Packet sizes for these applications are typically small, i.e. only a few bytes need to be transmitted.

URLLC features several system design trade-offs which deserve a study on their own. In what follows, we zoom in on some of these tradeoffs:

%Moreover, a thorough study smashing idealized assumptions and performance sensitivity to channel reciprocity, quasi-static channels and independence in fading realizations is needed:

\noindent \textbf {Finite versus large blocklength}:  In low-latency applications with small blocklength, there is always a probability that transmissions fail due to noise, deep fading, collision, interference, and so forth. In this case, the maximum coding rate $R_i(n,\epsilon)=k/n$ is lower than the Shannon rate, when transmitting  $k$ information bits using coded packets spanning $n$ channel uses.  Furthermore, for high reliability, data must be encoded at a rate which is significantly lower than the Shannon capacity\footnote{If the blocklength is large, no errors occur and the achievable rate is equal to the well-known Shannon capacity, $R_i=\log_2(1+{\rm SNR})$ for AWGN channels.}.
A number of works have shown that the Shannon capacity model significantly overestimates the delay performance for such applications, which would lead to insufficient resource allocations \cite{ArnauK17aa}.  Despite a huge interest in the field \cite{5452208,DurisiKP16}, a solid theoretical framework for modeling the performance of such systems due to short time spans,  finite blocklength and interference is needed.

\noindent \textbf {Spectral efficiency versus latency}: achieving low latency incurs a spectral efficiency  penalty (due to HARQ and shorter TTI). A system characterization of spectral efficiency versus latency in a multiple access and  broadcast system taking into account: (a) bursty packet arrivals, (b) mixture of low latency and delay  tolerant traffic, and (c) channel fading and multipath is missing. Low latency transmission on the uplink can be achieved with single shot slotted Aloha  type transmission strategy where the device sends  data immediately without incurring the delay  associated with making a request and receiving a scheduling grant.

\noindent \textbf {Device energy consumption versus latency}: A  fundamental tradeoff that needs to be characterized is the relationship between device energy consumption and latency. In wireless communications, devices need to be in sleep or deep sleep mode when they are not transmitting or  receiving to extend battery life. Since applications from the network may send packets to the device, the device needs to wake up periodically to check if packets are waiting. The frequency with which the device checks for incoming packets determines latency of the packets as well as the energy consumption. The more frequent the lower the latency but higher the energy consumption. Similarly, when reliability is deteriorated retransmissions are required, which means more energy consumption is needed to guarantee a target reliability.

\noindent \textbf {Energy expenditures versus reliability}:   higher reliability requires having several low-power transmissions instead of one high-reliable and high-power transmission as shown in \cite{7403840}. However, this depends on the diversity order that can be achieved and whether independent or correlated fading is considered.

\noindent \textbf {Reliability versus latency and rate}: generally speaking, higher reliability requires higher latencies due to retransmissions, but there could also be cases where both are optimized. In terms of data rates, it was shown in \cite{ParkP17} that guaranteeing higher rates incur lower reliability and vice-versa.

\noindent \textbf {SNR versus diversity}: fundamental questions include: how does the SNR requirement decrease as a function of network nodes and diversity order? Does the use of higher frequency bands help provide more/less diversity? How much SNR is needed to compensate for time-varying channels, bad fading events, etc. Besides, how does reliability scale with the number of links/nodes?

\noindent \textbf {Short/long TTI  versus control overhead}:  TTI duration should  be adjusted according to   user-specific radio channel conditions and quality of service requirements to compensate for the control overhead. As shown in \cite{7600901},  as the load increases the system must gradually increase the TTI size (and consequently the spectral efficiency) to cope with  non-negligible queuing delay, particularly for the tail of the latency distribution. Hence using different TTI sizes to achieve low latency, depending on the offered load and the percentile of interest is needed.

\noindent \textbf {Open  versus closed loop}: For closed loop if more resources are used for channel training and estimation, more accurate CSI is obtained albeit low data resources. For open loop a simple broadcast to all nodes may be sufficient but requires more downlink resources. In this case building diversity by means of relay nodes or distributed antennas is recommended. This problem needs to be revisited in light of short packet transmissions.

\noindent \textbf {User density versus dimensions (antennas, bandwidth and blocklength)}:  Classical information theory rooted in infinite coding blocklength assumes a fixed (and  small) number of users, where fundamental limits such as the coding blocklength $n \rightarrow \infty$ are studied. In  the large-scale  multiuser scenario, $n \rightarrow \infty$ is taken  before the number of users $k \rightarrow \infty$. In massive machine type communication, a massive number of  devices with sporadic traffic  need to share the spectrum in a given area which means $k>n$. In this case, allowing $n \rightarrow \infty$ while fixing $k$ may be inaccurate and provides little insight.  This warrants a  rethinking of the assumption of fixed population of full buffer users. A step towards this vision is \emph{many-user} information theory  where the number of users  increases without bound with the blocklength, as proposed in \cite{ChenCG17}.

\section{Tools and Methodologies for URLLC}

As alluded to earlier, URLLC mandates a departure from expected utility-based approaches relying on average quantities. Instead, a holistic framework which takes into account end-to-end delay, reliability, packet size, network architecture/topology, scalability and decision-making under uncertainty is needed.  In addition, fundamental system design and algorithm principles central to URLLC are at stake.  Next, following-up on the breakdown in Figure 1, we   identify a (non-exhaustive) set of tools and methodologies which serve this purpose.
\subsection{RISK}
\subsubsection{\textbf{Risk-sensitive learning and control}}

The notion of ``risk'' is defined as the chance of a \emph{huge loss} occurring with \emph{very low} probability. In this case instead of maximizing the expected payoff (or utility), the goal is to mitigate the risk of the huge loss. While reinforcement learning aims at maximizing the expected utility of an agent (i.e., a transmitting node), risk-sensitive learning is based on the fact that the utility is modified so as to incorporate the risk (e.g., variance, skewness, and other higher order statistics). This is done  by exponentiating  the agent's cost function before taking the expectation, yielding higher order moments. More concretely, the utility function of an agent $i$ is given by:
\begin{equation}
r_{i,\mu_i}=\frac{1}{\mu_i} \log \Big[\mathbb{E}_{\underline{x}}(e^{\mu_iu_i(\underline{x}_i)})\Big],
\end{equation}
where $\mu_i$ is the risk-sensitivity index and $\underline{x}_i$ is the agent's transmission strategy.  By doing a Taylor approximation around $\mu_i=0$ we get:
\begin{equation}
r_{i,\mu_i}\approx \mathbb{E}[u_i]+\frac{\mu_i}{2}{\rm var}(u_i)+ {\rm O}(\mu_i).
\end{equation}
Moreover,
\begin{equation}
\lim_{\mu_i \rightarrow 0} \frac{r_{i,\mu_i}-1}{\mu_i}=\lim_{\mu_i \rightarrow 0} \frac{1}{\mu_i} \Big[\mathbb{E}_{\underline{x}_i}(e^{\mu_iu_i(\underline{x}_i)})-1\Big]=\mathbb{E}_{x_i}(u_{i}).
\end{equation}
In risk-sensitive reinforcement learning, every agent needs to first estimate its own utility function $\hat{r}_{i,\mu_i}$ over time based on a (possibly delayed or imperfect) feedback before updating its transmission probability distribution $\underline{x}_i$. The utility estimation of agent $i$ when choosing strategy $\underline{x}_i$ is typically given by:

\begin{equation}
\hat{r}_{i,t+1}(x_i)=\hat{r}_{i,t}(x_i)+\lambda_t \Big(\frac{e^{\mu_iu_{i,t+1}}-1}{\mu_i}-\hat{r}_{i,t}(x_i)\Big),
\end{equation}
where $\lambda_t$ is a learning parameter. The application of risk-sensitive learning in the context of millimeter-wave communication is given in Section VIII.A.

\subsubsection{\textbf{Mathematical finance and portfolio optimization}}

Financial engineering and electrical engineering are seemingly different areas that share strong underlying connections. Both areas rely on statistical analysis and modeling of systems and the underlying time series. Inspired from the notion of risk in mathematical finance, we examine various risk measures, such as the value-at-risk (VaR), conditional VaR (CVaR), entropic VaR (EVaR) and mean-variance.
\begin{itemize}
\item	\textbf{Value-at-Risk ($\rm VaR$)}:  Initially proposed by J.-P. Morgan, $\rm VaR$ was developed in response to financial disasters of the 1990s and played a vital role in market risk management. By definition $\rm VaR$ is the worst loss over a target horizon with a given level of confidence such that for $0<\alpha\leq 1$:
\begin{equation}
{\rm VaR}_{1-\alpha}(X)=\inf_t\Big\{t: {\rm Prob}(X\leq t) \geq 1-\alpha\Big\}
\end{equation}
which can also be expressed as: ${\rm VaR}_{\alpha}(X)=F_{X}^{-1}(1-\alpha)$.
\item	\textbf{Conditional VaR}: CVaR measures the expected loss in the right tail  given a particular threshold has been crossed.  The CVaR is defined as the conditional mean value of a random variable exceeding a particular percentile. This precisely measures the risky realizations, as opposed to the variance that simply measures how spread the distribution is. Moreover CVaR overcomes the caveat of VaR due to the lack of control of the losses incurred beyond the threshold. Formally speaking, it holds that:
\begin{align}
{\rm CVaR}_{1-\alpha}(X)&=\inf_t\Big\{t+\frac{1}{\alpha} \mathbb{E}[(X-t)^+]\Big\}\notag
\\&=\frac{1}{\alpha} \int_{0}^{\alpha}{\rm VaR}_{1-t}(X)dt =\mathbb{E}(X | X> {\rm VaR}_{1-\alpha}).
\end{align}
\item	\textbf{Entropic VaR (EVaR)}: {$\rm EVaR$} is the tightest upper bound one can find using the Chernoff inequality for the ${\rm VaR}$, where for all $z \geq 0$:

\begin{equation}
{\rm Prob}(X \geq a) \leq e^{-az}M_X(z)
\end{equation}
 $M_X(z)=\mathbb{E}(e^{zX})$ is the moment generating function (MGF) of  random variable $X$.

 \begin{equation}
 {\rm EVaR}_{1-\alpha}(X)= \inf_{z>0} z^{-1} \log (M_X(z))/\alpha
\end{equation}

By solving the equation $e^{-za}M_X(z)=\alpha$ with respect to $a$ for $\alpha \in [0,1]$, we get
 \begin{equation}
 a_X(\alpha,z)=z^{-1}\log (M_X(z))/\alpha.
\end{equation}

$\rm EVaR$ is an upper bound for the CVaR and its dual representation is related to the Kullback-Leibler divergence \cite{Ahmadi}. Moreover, we have that:

\begin{equation}
\log \mathbb{E}_P(e^X)= \sup_{Q<<P} \mathbb{E}_Q(X)-D_{KL}(Q || P )
\end{equation}

%In addition, it can be shown that \cite{Ahmadi}: ${\rm EVaR}_{1-\alpha}(X)= \sup_{Q } \mathbb{E}_Q(X).$

\noindent \textbf{Remark}: Interestingly, we note that VaR and CVaR are related to the Pickands-Balkema-de Haan theorem of EVT, i.e., Theorem \ref{Thm: GPD}. To illustrate that, we denote ${\rm VaR}_{1-\alpha}(X)=d$. Since ${\rm CVaR}_{1-\alpha}(X)$ calculates the mean of $X$ conditioned on $X> {\rm VaR}_{1-\alpha}$, we have that:
\begin{equation}\label{Eq: VaR_EVT}
{\rm CVaR}_{1-\alpha}(X)=\mathbb{E}(X | X> {\rm VaR}_{1-\alpha})=\mathbb{E}(Y+d | X> d)=\mathbb{E}(Y| X> d)+d,
\end{equation}
where $Y|_{X>d}=X-d$. Letting $\alpha\to 0$ and as per Theorem \ref{Thm: GPD}, $Y$ can be approximated by a  generalized Pareto distribution random variable whose mean is equal to ${\rm CVaR}_{1-\alpha}(X)-{\rm VaR}_{1-\alpha}(X)$.

\item	\textbf{Markowitz's mean-variance (MV)}: MV is one of the most popular risk models in modern finance (also referred to as Markowitz's risk-return model),  in which the value of an investment is modeled as a tradeoff between expected payoff (mean return) and variability of the payoff (risk). In  a  learning context, this entails learning the variance of the payoff from the feedback as follows.  First, the variance estimator of agent $i$ at time $t$ is given by:
\begin{equation}
\hat{v}_{i,t}=\frac{1}{t}\sum_{t'=1}^{t}(r_{i,t'}-\hat{r}_{i,t'})^2,
\end{equation}
after which the variance is estimated as:
\begin{equation}
\hat{v}_{i,t+1}=\hat{v}_{i,t}+\beta_t \Big( (r_{i,t'}-\hat{r}_{i,t'})^2-\hat{v}_{i,t}\Big).
\end{equation}
\end{itemize}
and $\beta_t$ is a learning parameter. Once the variance is estimated,  principles of reinforcement learning can be readily applied \cite{6542770}.

\noindent \noindent \textbf{Remark}: Risk-sensitivity is also a common theme in prospect theory (PU), in which the notion of risk arises through nonlinear distortions of the value function and probabilities. PU dates back to the Nobel prize work of Kahnemann and Tversky \cite{kahneman2013prospect}, whose postulates depart from expected utility theory and risk-neutrality. Instead, it is based on two factors, namely the distortion of probabilities in the probability weighting function and the shape in the value function. This is motivated by the fact that people put too much weight on small probabilities and too little weight on large probabilities. Mathematically, this is captured by a  value function and probability weighting function whose shape is determined by a set of parameters. The application of prospect theory in wireless communication can be found in cognitive radios \cite{YangPMSGS15} and smart grid \cite{RahiSSMP16}.
\subsection{TAIL}

\subsubsection{\textbf{Extreme value theory (EVT)}}

Latency and reliability are fundamentally about “taming the tails” and going beyond “central-limit theorems”. In this regard, EVT provides a powerful and robust framework to fully characterize the probability distributions of extreme events and extreme tails of distributions.  EVT has been developed during the twentieth century and is now a well-established tool to study extreme deviations from the average of a measured phenomenon. EVT has found many applications in oceanography, hydrology, pollution studies, meteorology, material strength, highway traffic and many others (for a comprehensive survey on EVT, see \cite{Coles01}). EVT is built around  the two  following Theorems:
%: the \emph{Fisher-Tippett-Gnedenko} theorem and the \emph{Pickands-Balkema-de Haan} theorem.

\begin{theorem}[{\bf Fisher-Tippett-Gnedenko theorem for block maxima} \cite{Coles01}]\label{Thm: GEV}
Given $n$ independent samples,  $X_1,\cdots,X_n$, from the random variable $X$,  define $M_n=\max\{X_1,\cdots,X_n\}$. As $n\to\infty$, we can approximate the cumulative distribution function (CDF)  of $M_n$ as
\begin{equation}
\label{Eq: GEV}
F_{M_n}(z)={\rm Prob}(M_n\leq z)
\approx G(z;\mu,\sigma,\xi)=
e^{-\big(1+\frac{\xi (z-\mu)}{\sigma}\big)^{-1/\xi}},
\end{equation}
where $G(z;\mu,\sigma,\xi)$, defined on $\{z\!\!:1+\xi (z-\mu)/\sigma>0\}$, is the generalized extreme value (GEV) distribution characterized by the location parameter $\mu\in\mathbb{R}$, the scale parameter $\sigma >0$, and the shape parameter $\xi \in\mathcal{R}$.
\end{theorem}

\begin{theorem}[{\bf Pickands-Balkema-de Haan theorem for exceedances over thresholds} \cite{Coles01}]\label{Thm: GPD}
Consider the distribution of X conditionally on exceeding some high threshold $d$. As the threshold $d$ closely approaches $F^{-1}_{X}(1)$, i.e., $d\to\sup\{x\!\!: F_{X}(x)<1\}$,  the conditional CDF of the excess value $Y=X-d>0$ is
\begin{equation}\label{Eq: GPD}
 F_{Y|X>d}(y)={\rm Prob}(X-d\leq y|X>d)\approx H(y;\tilde{\sigma},\xi)=
 1-\bigg(1+\frac{\xi y}{\tilde{\sigma}}\bigg)^{-1/\xi},
\end{equation}
where $H(y;\tilde{\sigma},\xi)$, defined on $\{y\!\!:1+\xi y/\tilde{\sigma}>0\}$, is the generalized Pareto distribution (GPD).
Moreover, the characteristics of the GPD depend on the scale parameter $\tilde{\sigma} >0$ and the shape parameter $\xi \in\mathcal{R}$. The location and scale parameters in \eqref{Eq: GEV} and \eqref{Eq: GPD} are related as per $\tilde{\sigma}=\sigma+\xi(d-\mu)$ while the shape parameters $\xi$ in both theorems are identical.
\end{theorem}

While Theorem \ref{Thm: GEV} focuses on the maximal value of a sequence of variables, Theorem \ref{Thm: GPD} aims at the values of a sequence above a given threshold. Both theorems asymptotically characterize the statistics of extreme events and provide a principled approach for the analysis of ultra-reliable communication, i.e., failures with extreme low probabilities. A direct application of EVT to mobile edge computing scenarios is found in Section VII.D. Other applications of EVT in vehicular communication can be found in \cite{LiuB18,FLBENNIS}.

\noindent \subsubsection{\textbf{Effective bandwidth}}
Effective bandwidth is a large-deviation type approximation defined as the minimal constant service rate needed to serve a random arrival under a queuing delay requirement \cite{515976}.  Let $a(t)$ and $q(t)$ be the number of arrivals at time $t$ and the number of users in the queue at time $t$, respectively. Assume the queue size is infinite and the server can serve $c(t)$ users per unit of time. $c(t)$ is referred to as server capacity at time $t$ and the queue is governed by the following equation:
 \begin{equation}
q(t+1)=\Big(q(t)+a(t+1)-c(t+1)\Big)^+
 \end{equation}
Moreover, let:
\begin{equation}
 \Lambda_A(\theta) = \lim_{t \rightarrow \infty} \frac{1}{t} \log \mathbb{E}[e^{\theta A(0,t)}]
 \end{equation}
 for all $\theta \in \mathbb{R}$, and $\Lambda_A(\theta)$ is differentiable.
 Let $\Lambda_A^*(\alpha)$ be the Legendre transform of  $\Lambda_A(\theta)$, i.e.,
 \begin{equation}
 \Lambda_A^*(\alpha)=\sup_{\theta} \big[\theta \alpha - \Lambda_A(\theta)\big].
  \end{equation}
Likewise let $\Lambda_C^*(\alpha)$ be the Legendre transform of $\Lambda_C(\theta)$. We seek the decay rate of the tail distribution of the stationary queue length. This is given in \cite{515976} which states that if there exists a unique $\theta^*>0$ such that
\begin{equation}
 \Lambda_A^*(\alpha)+ \Lambda_C^*(-\alpha)=0,
  \end{equation}

\noindent then it holds that:
\begin{equation}
\lim_{x \rightarrow \infty} \frac{\log{\rm Pr}(q(\infty  \geq x))}{x}=-\theta^*.
 \end{equation}

\noindent  In particular, for fixed capacity $c(t)=c$ for all $t$, we have that:
\begin{equation}
\frac{\Lambda_A(\theta^*)}{\theta^*}=c
 \end{equation}

\noindent $\frac{\Lambda_A(\theta^*)}{\theta^*}$ is called the effective bandwidth of the arrival process subject to the condition that the tail distribution of the queue length has the decay rate $\theta^*$.

\noindent \textbf{Remark}:  since the distribution of queuing delay is obtained based on large deviation principles, the effective bandwidth can be used for constant arrival rates, when the delay bound is large and the delay violation probability is small. This raises the question on the usefulness and correctness of using the effective bandwidth in solving  problems dealing with finite packet/queue lengths and very low latencies.

\subsubsection{\textbf{Stochastic network calculus (SNC)}}

SNC considers queuing systems and networks of systems with stochastic arrival, departure, and service processes, where the bivariate functions $A(\tau,t)$, $D(\tau,t)$ and $S(\tau,t)$ for any $0\leq \tau \leq t$ denote the cumulative arrivals, departures and service of the system, respectively, in the interval $[\tau,t)$. The analysis of queuing systems is done through simple linear input-output relations. In the bit domain it is based on a $(\min,+)$ dioid algebra where the standard addition is replaced by the minimum (or infimum) and the standard
multiplication replaced by addition. Similar to the convolution and deconvolution in standard algebra, there are definitions for convolution and deconvolution operators in the $(\min,+)$ algebra and  the convolution and deconvolution operators in $(\min,+)$-algebra are often used for performance evaluation. Finally, observing that the bit and SNR domains are linked by the exponential function, arrival and departure processes are transferred from the bit to the SNR domain. Then  backlog and delay bounds are derived in the  transfer domain using the $(\min,\times)$ algebra before going back to the bit domain to obtain the desired performance bounds.

%via the Mellin transform or moment generating function (MGF) is converting the accumulatively transmitted data and arrived data from the bit domain to the SNR domain, where an upper bound of delay bound violation probability can be obtained.

Define the cumulative arrival, service and departure processes as:

 \begin{equation}
A(\tau,t)=\sum_{i=\tau}^{t-1}a_i, \quad S(\tau,t)=\sum_{i=\tau}^{t-1}s_i, \quad D(\tau,t)=\sum_{i=\tau}^{t-1}d_i
\end{equation}

\noindent  The backlog at time t is given by $B(t)=A(0,t)-D(0,t)$. Moreover,  the delay $W(t)$ at time $t$, i.e. the number of slots it takes for an information bit arriving at time t to be received at the destination, is $W(t)=\inf \Big\{u \geq 0: A(0,t)/D(0,t+u) \leq 1\Big\}$ and the delay violation probability is given by

 \begin{equation}
\Lambda(w,t)={\rm Prob}(W(t) > w)
\end{equation}

SNC allows to obtain bounds on the delay violation probability based on simple statistical characterizations of the arrival and service processes in terms of their Mellin transforms. First by converting the cumulative processes in the bit domain through the exponential function, the corresponding processes in the SNR domain are: $A(\tau,t)=e^{A(\tau,t)}, \quad S(\tau,t)=e^{S(\tau,t)}, \quad D(\tau,t)=e^{D(\tau,t)}$.  From these definitions, an upper bound on the delay violation probability can be computed by means of the Mellin transforms of $A(\tau,t)$ and $S(\tau,t)$:

\begin{equation}
p_v(w) = \inf_{s>0}\{K(s,-w)\} \geq   \Lambda(w)={\rm Prob}(W > w)
\end{equation}
where $K(s,-w)$ is  the so-called steady-state kernel, defined as

 \begin{equation}
\mathcal{K}(s,-w)= \lim_{t \rightarrow \infty} \sum_{u=0}^\tau \mathcal{M}_\mathcal{A}(1+s,t+w,t)\cdot \mathcal{M}_\mathcal{S}(1-s,t+w,\tau)
\end{equation}

\noindent and $\mathcal{M}_X(s)=\mathbb{E}[X^{s-1}]$ denotes the Mellin\footnote{By setting $s=\theta+1$, we obtain the effective bandwidth and MGF-based network calculus.} transform of a nonnegative random variable $X$, for any $s$.

Analogous to \cite{mellin} we consider $(\sigma(s), \rho(s))$-bounded arrivals where the log-MGF of the cumulative arrivals in the bit domain is bounded by

 \begin{equation}
\frac{1}{s}\log \mathbb{E}[e^{sA(\tau,t)}] \leq \rho(s).(t-\tau)+\sigma(s).
\end{equation}
This characterization can be viewed as a probabilistic extension of a traffic flow that is deterministically regulated by a token bucket with rate $\rho$ and burst size $\sigma$.
To simplify the notation, we restrict the following analysis to values $(\sigma, \rho)$ that are independent of $s$, which is true for constant arrivals. Subsequently, the Mellin transform of the SNR-domain arrival process can be upper-bounded by:

 \begin{equation}
\mathcal{M}_\mathcal{A}(s,\tau,t)   = \mathbb{E}[A(\tau,t)^{s-1}] \leq e^{(s-1)(\rho.(t-\tau)+\sigma)}
\end{equation}

Assuming the cumulative arrival process in SNR domain to have stationary and independent increments, the steady-state kernel for a fading wireless channel is given by:
\begin{equation*} \mathcal{K}(s,\ -w)=\frac{\mathcal{M}_{g(\gamma)}^{w}(1-s)}{1-\mathcal{M}_{\alpha}(1+s)\mathcal{M}_{g(\gamma)}(1-s)} \tag{7} \end{equation*}

\noindent  for any $s > 0$, under the stability condition $\mathcal{M}_{\alpha}(1+s)\mathcal{M}_{g(\gamma)}(1-s) < 1.$  The delay bound (24) thus reduces to

\begin{equation}
p_v(w) = \inf_{s>0}\frac{\mathcal{M}_{g(\gamma)}^{w}(1-s)}{1-\mathcal{M}_{\alpha}(1+s)\mathcal{M}_{g(\gamma)}(1-s)}
\end{equation}

\subsubsection{\textbf{Meta distribution}}
The meta distribution is a fine-grained key performance metric of wireless systems, first introduced in \cite{Haenggi16} which provides a mathematical foundation for questions of network densification under strict reliability constraints. As such, the meta distribution is a much sharper and refined metric than the standard success probability, which is easily obtained as the average over the meta distribution. By definition, the meta distribution $\bar {F}_{P_{\mathrm{ s}}}(x)$ is the complementary cumulative distribution function (CCDF) of the random variable

\begin{equation}
P_{\mathrm{ s}}(\theta )\triangleq \mathbb {P}({\rm SIR_{0}}>\theta \mid \Phi)=\mathbb{E}_h [1_{\{ {\rm SIR_{0}}>\theta \}}]
 \end{equation}
which is the CCDF of the conditional signal-to-interference ratio (SIR) of the typical user ($'0'$) given the points processes $\Phi$ and conditioned on the desired transmitter to be active. The meta distribution is formally given by \cite{5757524}:

\begin{equation}
\bar {F}_{P_{\mathrm{ s}}}(x)\triangleq \mathbb {P}^{0}(P_{\mathrm{ s}}(\theta )>x),\quad \theta \in \mathbb{R}^+, \quad x\in \left [{ 0,1 }\right ].
\end{equation}
 $\mathbb {P}^{0}$ is the Palm measure. Interestingly, the moments $M_b$  reveal interesting properties of the meta distribution, in which
\begin{equation}
M_1=\int_{0}^{1}\bar {F}(\theta, x)dx \quad {\rm and} \quad {\rm var} (P_{\mathrm{ s}}(\theta))= M_2-M_1^2.
\end{equation}
is the standard success probability and variance, respectively.

Since all point processes in the model are ergodic, the meta distribution can be interpreted as the fraction of the active links whose conditional success probabilities are greater than $x$.   A  simple  approach to calculate the meta distribution is  to approximate it with the beta distribution, which requires only the first and second moments. Recent applications of the Meta distribution can be found in \cite{ParkP17} for industrial automation,   \cite{AbdullaW17} in the context of V2V communication, and millimeter-wave device-to-device networks in \cite{DengH17}.

\subsection{SCALE}

\subsubsection{\textbf{Statistical physics}} \label{sec:mean_field_sumudu}

Current wireless systems can support tens to hundreds of nodes with latency constraints on the order of seconds and with moderate reliability. Nonetheless, these do not scale well to systems with thousands to millions of nodes as envisaged in massive MTC or ultra-dense networks. Solving resource allocation problems in  large-scale network deployments is intractable and requires cumbersome Monte-Carlo simulations lacking  fundamental insight. Here, cutting-edge methods from statistical physics such as the replica method and cavity methods, provide fundamental insights and guidelines for the design of massive and ultra-reliable networks.

%********************************************
\newcommand{\set}[1]{\mathcal{#1}}%{\mathpzc{#1}}
\newcommand{\vect}[1]{\boldsymbol{#1}}
\newcommand{\size}[1]{|\set{#1}|}
\newcommand{\directdelta}{\delta}
\newcommand{\expect}{\mathbbm{E}}
\newcommand{\partialX}[2]{\frac{\partial{#1}}{\partial{#2}}}
\newcommand{\optimal}{^\star}
\newcommand{\hamiltonion}{\mathcal{H}}%{\aleph}

\newcommand{\connectivity}{x}
\newcommand{\connectivityvect}{\vect{\connectivity}}
\newcommand{\energy}{\bar{E}}
\newcommand{\partitionSum}{Z}
\newcommand{\temperature}{\beta}
\newcommand{\replica}{n}
\newcommand{\cost}{\Phi}
\newcommand{\costOptimal}{\cost\optimal(\channel)}
\newcommand{\channelsingle}{h}
\newcommand{\channel}{\vect{\channelsingle}}

\newcommand{\ACT}{A}
\newcommand{\actionsingle}{a}
\newcommand{\action}{\vect{\actionsingle}}
\newcommand{\actionOthers}[1]{\action_{-{#1}}}
\newcommand{\player}{b}
\newcommand{\PLAYER}{B}
\newcommand{\playerset}{\set{\PLAYER}}
\newcommand{\actionspace}{\set{\ACT}}
\newcommand{\vectx}{\vect{x}}
\newcommand{\vectxx}{\breve{\vect{x}}}
\newcommand{\vecty}{\vect{y}}
\newcommand{\vectY}{\vect{Y}}
\newcommand{\state}{\vectx}
\newcommand{\statex}{\vectxx}
\newcommand{\actionx}{\breve{\action}}
\newcommand{\statesingle}{x}
\newcommand{\statevector}{\vect{X}}
\newcommand{\stateset}{\set{X}}
\newcommand{\massdistribution}{\rho}
\newcommand{\utility}{u}

\newcommand{\sinr}{\gamma}
\newcommand{\BS}{B}
\newcommand{\UE}{U}
\newcommand{\ratio}{\zeta}
\newcommand{\countBS}{v_b}
\newcommand{\countUE}{\nu_u}
\newcommand{\powerMxBS}{\vartheta}
\newcommand{\powerMxUE}{\varphi}
%*************************************************

Interactions between particles (e.g. atoms, gases or neurons) is a common theme in the statistical physics literature~\cite{jnl:hubbard59,jnl:nieuwenhuizen98,jnl:dellerba12}.
Here, instead of analyzing the  microscopic network state,  one is concerned with the macroscopic state requiring only a few parameters.
By invoking this analogy and  modeling network elements as particles, a network cost function $\cost(\state,\channel)$ which captures the interactions between network elements is referred to as the Hamiltonian $\hamiltonion(\state\in\stateset)$ over the state space $\stateset$ and   $\channel$.

One of the underlying principles of statistical physics is to find the lowest energy of a system  with the aid of the \emph{partition sum} $\partitionSum = \sum_{\state\in\stateset} e^{-\temperature \hamiltonion(\state)}$, summing over all the states of the Boltzmann factors $e^{-\temperature\hamiltonion(\connectivityvect)}$ at a given fictitious temperature $\frac{1}{\temperature}\to 0$.
By adopting the concept of quenched disorder%
\footnote{
	The quenched disorder of a system exists when the randomness of the system characteristics is time invariant.
}%
, the network can be replicated to solve the interactions of infinitely large number of network elements, in which the ground state over  $\expect_{\channel}[\costOptimal]$ is found via the \emph{replica method}.
Here, $\costOptimal$ is the optimal  network cost with the exact knowledge of the randomness $\channel$ (e.g. channel and queue states).
The replica method in statistical mechanics refers to the idea of computing moments of vanishing order, in which the word ``replica" comes from the presence of $n$ copies of the vector of configurations in the calculation of the partition function.
The \emph{quenched average free energy} $\energy$, which is calculated by employing the replica trick on the partition sum ($\replica$ times replicating $\partitionSum$) is given by,
\begin{equation}\label{eqn:quenched_avg_energy}
\energy = \lim_{\replica\to 0} \frac{\energy_{\replica}}{\replica} = \lim_{\replica\to 0} \frac{-\ln\overline{\partitionSum^\replica}}{\temperature\replica} = \expect_{\channel}[\costOptimal],
\end{equation}
where $\energy_{\replica}=\frac{-\ln\overline{\partitionSum^\replica}}{\temperature}$ is the \emph{replicated free energy}.
$\overline{\partitionSum^\replica}$ is the quenched average of the $\replica$-th replica of the partition sum.
Deriving a closed form expression for the replicated free energy $\energy$ is the key to analyze the performance of dense network deployments.

\subsubsection{\textbf{Mean field game theory}} \label{sec:mean_field_sumudu}
When a large number of  wireless nodes compete over limited resources (as in massive machine type communication or ultra-dense networks), the framework of mean-field games (MFGs) is instrumental  in  studying multi-agent resource allocation problems without resorting to time-consuming Monte-Carlo simulations \cite{book:gueant11,book:caines14,KimPBKD17}.  In a game with very large number of agents, the impact of every agent on the other agents' utility functions is infinitesimal and the equilibrium/game is dominated by a nontrivial proportion of the population (called \emph{mean field}).  The game can therefore be analyzed at a macrolevel using mean field theory and fundamental limits of the network can be unraveled.  Applications of MFGs  are  of utmost importance, especially when \emph{scalability} and \emph{complexity} matter, such as when optimizing autonomous vehicles' trajectories or UAV platooning, distributed machine learning and many others.

Consider a state distribution $\vect{\massdistribution}^{\size{\PLAYER}}(t) = [ \massdistribution^{\size{\PLAYER}}\big(t,\state') ]_{\state'\in\set{X}}$ of set of players $\playerset$ where $\massdistribution^{\size{\PLAYER}}\big(t,\state') =  \frac{1}{\size{\PLAYER}} \sum_{\player=1}^{\size{\PLAYER}} \directdelta\big(\state_\player(t)=\state'\big)$
represents the fraction of players at each state $\state'$ in state space $\stateset$ with $\directdelta(\cdot)$ being the \emph{Dirac delta} function.
In the limit of $\size{\PLAYER}\to\infty$, the goal in mean field (MF) is to seek an equilibrium with,
\begin{equation}\label{eqn_SOA:mass_distribution_infinite}
\vect{\massdistribution}\big(t) =  \bigg[ \lim\limits_{\size{\PLAYER}\to\infty} \frac{1}{\size{\PLAYER}} \textstyle  \sum\limits_{\player=1}^{\size{\PLAYER}} \directdelta\big(\state_\player(t)=\state'\big) \bigg]_{\state'\in\stateset},
\end{equation}
which is known as the \emph{MF distribution}.
Assuming a  homogeneous control policy over all players, i.e. $\action_\player\big(t,\state(t) \big) = \action\big(t,\state_\player(t),\massdistribution^{\size{\PLAYER}}(t) \big)\in\actionspace$ for all players ${\player\in\playerset}$, their interactions can be represented as a game consisting of a set of generic players with the state $\statex(t)$ and action $\actionx(t)$ at time $t$.
Here, the goal of a generic player is to maximize its  own utility $\utility\big(t,\statex(t)\big)
= \textstyle \expect \big[ \int_t^T  f\big(\tau,\statex(\tau),\action(\tau),\massdistribution(\tau)\big) d\tau \big]$ over the actions $\actionx\big(t,\statex(t)\big)\in\breve{\actionspace}\big(t,\statex(t)\big)$ and time period $t\in[0,T]$ under the dynamics of states $ d\statex(t) = \breve{\statesingle}_t dt + \breve{\state}_z d\vect{z}(t)$ consist of both time dependent $\breve{\statesingle}_t$ and random $\breve{\state}_z$ components.
Remarkably,  the presence of a very large number of players  leads to a continuum allowing to obtain a solution for the above problem, a \emph{mean-field equilibrium}, using only two coupled partial differential equations (PDEs), Hamilton-Jacobi-Bellman (HJB) and Fokker-Planck-Kolmogorov (FPK) equations, respectively.
\begin{equation}\label{eqn_SOA:MFG_PDEs}
\begin{cases}
\partialX{}{t}[\utility\big(t,\statex(t)\big)] + \max\limits_{\actionx(t,\statex(t))} \bigg[ \breve{\statesingle}_t \partialX{}{\statex}[\utility\big(t,\statex(t)\big)] + f(t,\statex(t),\actionx(t)\big)
+ \frac{1}{2}\text{tr}\Big( \statex_z^2\frac{\partial^2}{\partial {\statex}^2}[\utility\big(t,\statex(t)\big)] \Big) \bigg]  = 0, \\
\partialX{}{t}\massdistribution\big(t,\statex(t)\big) + \partialX{}{\statex} \Big[ \partialX{}{\statex}[\utility\big(t,\statex(t)\big)]\massdistribution\big(t,\statex(t)\big) \Big] - \frac{1}{2}\text{tr}\Big( \statex_z^2\frac{\partial^2}{\partial {\statex}^2}[\massdistribution\big(t,\statex(t)\big)] \Big) = 0,
\end{cases}
\end{equation}
Furthermore, the optimal strategy is  given by:
\begin{equation}\label{eqn_SOA:optimal_strategy_MFG}
\textstyle \actionx\optimal\big(t,\statex(t)\big) = \argmax\limits_{\actionx(t,\statex(t))} \bigg[ \breve{\statesingle}_t \frac{\partial}{\partial x}[\utility\big(t,\statex(t)\big)] + f(t)
\textstyle  + \frac{1}{2}\text{tr}\Big( \statex_z^2\frac{\partial^2}{\partial x^2}[\utility\big(t,\statex(t)\big)] \Big) \bigg].
\end{equation}
which yields the optimal control of a generic player.

%\subsection{Beyond Shannon Entropy \textbf{(Mehdi)}}
%
%-->  TBC
\section{Case Studies}

To illustrate the usefulness and effectiveness of the URLLC methodologies, we focus our attention on four use cases pertaining to different verticals alongside their specific  requirements. While each of these use cases has distinct features, there are certain fundamental needs and principles that are common to all these applications, all stemming from the stringent requirements of ultra-low latency and high reliability. More specifically, first we will show how principles of risk-sensitive reinforcement learning can be applied to the problem of reliable millimeter-wave communication; second, in a virtual reality scenario  multi-connectivity  and proactive computing are shown to provide higher reliability and lower latency gains; third, extreme  value theory is shown to provide a principled and elegant tail-centric framework in a mobile edge computing scenario; fourth, tools from statistical physics are used to solve a user association problem in an ultra-dense network deployment scenario.

\subsection{Ultra-reliable millimeter-wave communication}
Owing to the vast chunks of spectrum in high-frequency bands, millimeter-wave communication  is  a key enabler for $5$G. However, operating at these frequencies suffers from high propagation loss  and link variability and susceptibility to blockage. In contrast to the classical network design based on average metrics (e.g., expected rate), we propose a risk-sensitive reinforcement learning-based framework to jointly optimize the beamwidth and transmit power of small cells (SCs) in a distributed manner, while taking into account the sensitivity of mmWave links due to blockage. The rationale for using risk-sensitive learning is to account for the fluctuations in the user's transmission rates while learning the best transmission strategy. To do that, every SC first estimates its own utility function based on user feedback and then updates its probability distribution over time for every selected strategy (i.e, transmit power and beamwidth) \cite{6542770,risksensitive}. Specifically, each SC adjusts its beamwidth from the range  $[0.2, 0.4]$ radian with a step size of $0.04$ radian. The transmit power level set of each SC is $\{ 21,23,25\}$ dBm. The number of transmit antennas and receive antennas at the SC and UE are set to $64$ and $4$, respectively. The blockage is modeled as a distance-dependent probability state where the channel is either line-of-sight (LOS) or non-LOS for urban environments at $28$ GHz with a $1$ GHz  system bandwidth. Numerical results are obtained via Monte-Carlo simulations over several random topologies.  Furthermore, we compare the proposed risk-sensitive learning (RSL) scheme with two baselines: (i) Classical learning (CSL) which refers to the learning framework in which the utility function only considers the mean value of the utility,  and (ii) Baseline 1 (BL1) where the SC selects the beamwidth with maximum transmit power.

\begin{figure*}[t]
 \centering
\begin{minipage}{0.45\linewidth}
     %\vspace{-1em}
	\includegraphics[width=1\columnwidth]{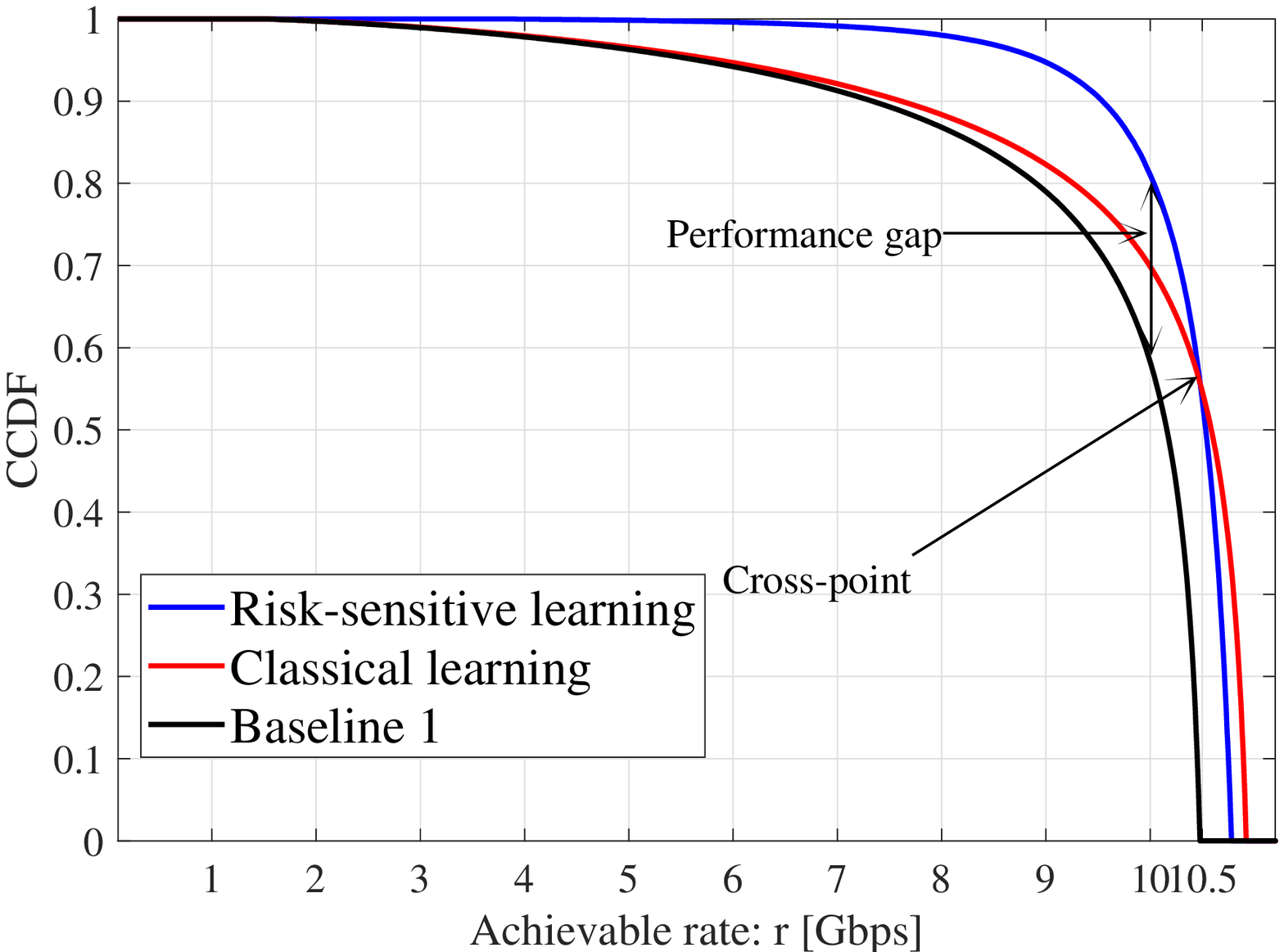}
	\caption{The tail distribution of the achievable rate, when the number of SCs is $B=24$.}
		\label{Fig_mmWave_CDF}
\end{minipage}
\hspace{1.5em}
\begin{minipage}{0.45\linewidth}
	 %\vspace{-1em}
	\includegraphics[width=1\columnwidth]{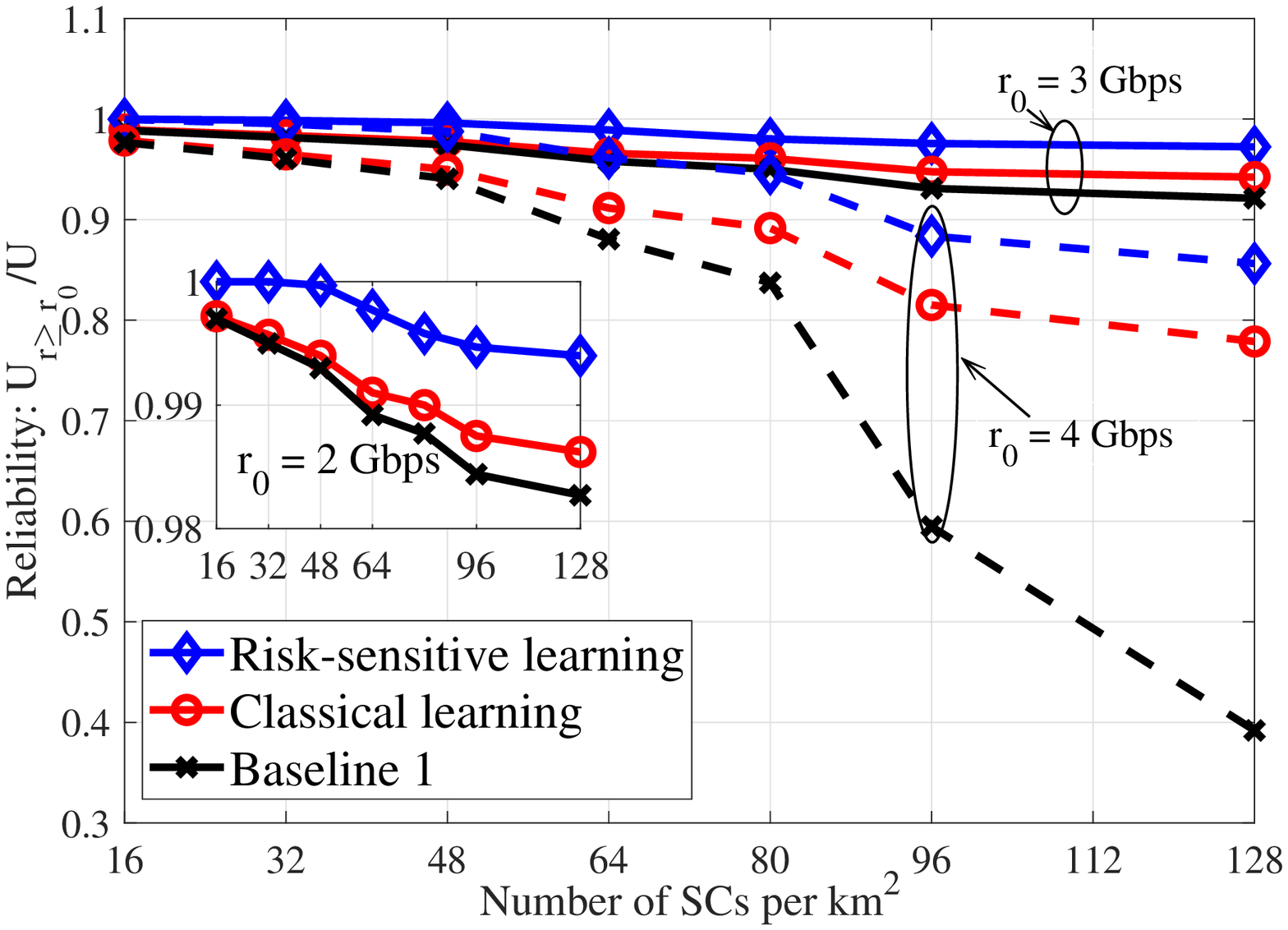}
	\caption{The tradeoff between reliability and network density.}
		\label{Fig_mmWave_Rel}
\end{minipage}

\end{figure*}

In Fig. \ref{Fig_mmWave_CDF}, we plot the  tail distribution, i.e., complementary cumulative distribution function (CCDF) of the achievable rate at $28$ GHz when the number of SCs is $24$ per $\text{km}^{2}$. The CCDF  captures the reliability defined as the probability that the achievable user rate is higher than a predefined target rate $r_{0}$, i.e, $\text{Pr\ensuremath{\left(\text{UR\ensuremath{\geq}r}_{0}\right)}}$. It is observed that the RSL scheme achieves a probability, $\text{Pr\ensuremath{\left(\text{UR\ensuremath{\geq}10 Gbps}\right)}}$, of more than $80\%$, whereas the baselines CSL and BL1 obtain less than $70\%$ and 60\%, respectively. However, at  lower rates (less than 2 Gbps) or very high rate ($10.62$-$11$ Gbps) shown by the cross-point, the RSL obtains a lower probability as compared to the baselines. This shows that the proposed solution provides a  user rate which is more concentrated around its median to provide uniform  service for all users. This can be seen from the user rate distribution in which RSL has a small variance of 0.5085, while the CSL and BL1 have a higher variance of 2.8678 and 2.8402, respectively.

\begin{figure}
\center
	 %\vspace{-1em}
	\includegraphics[width=0.5\columnwidth]{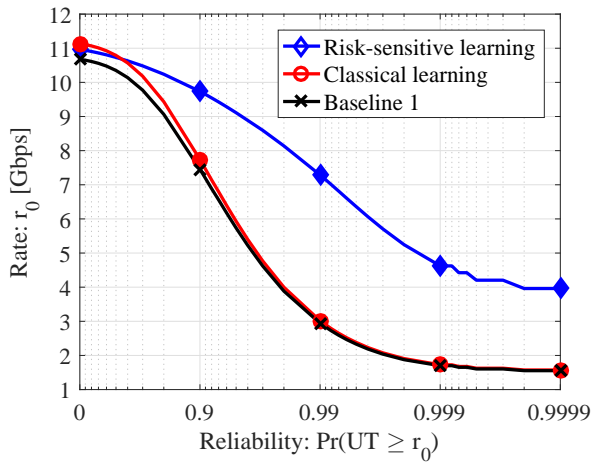}
	\caption{Rate-reliability tradeoff curves for the proposed risk-sensitive approach and baselines.}
		\label{Fig_mmWave_Rel1}
\end{figure}

Fig. \ref{Fig_mmWave_Rel} plots the impact of network density on the reliability, defined as the fraction of UEs who achieve a given target rate $r_{0}$, i.e., $\frac{U_{r>r_{0}}}{U}$. Here, the number of SCs varies from $16$ to $128$ per $\text{km}^{2}$. It is shown that for given target rates of $2$, $3$, and $4$ Gbps, RSL guarantees higher reliability as compared to the baselines. Moreover, the higher the target rate, the bigger the performance gap between the proposed algorithm and the baselines. Additionally, a linear increase in network density is shown to decrease reliability when density increases from $16$ to $96$, and the fraction of users that achieve $4$ Gbps of the RSL, CSL, and BL1 is reduced by $11.61\%$, $16.72\%$, and $39.11\%$, respectively. The rate reliability tradeoff is further corroborated in Fig. \ref{Fig_mmWave_Rel1}.

\subsection{Virtual reality (VR)}

VR is a  use case where ultra-reliable and low-latency communication plays an important role, due to the fact that the human eye needs to experience accurate and smooth movements with low ($<20$~ms) motion-to-photon (MTP) latency to avoid motion sickness. In our scenario, multiple players coexist in a gaming arcade and engage in a VR interactive gaming experience. VR head-mounted displays (HMDs) are connected via  mmwave wireless connections to multiple servers  operating in the same mmwave frequency band and are equipped with edge computing servers and storage units. To minimize the VR service latency, players offload their computing tasks which consist of  rendering high-definition video frames to the edge servers over mmWave links. First, players send their tracking data, consisting of their poses (location and rotation  coordinates) and game play data, in the uplink to an edge server. The edge server renders the corresponding player's frame and transmits it in the downlink. Since edge servers are typically equipped with high computation power graphical processing units (GPUs), computing latency is minimized as compared to local computing in the player's HMD. In addition to minimizing computing latency, reliable and low latency communication is needed to minimize the over-the-air communication latency.

\noindent In view of this, we propose a proactive computing and \emph{multi-connectivity (MC)} solution, which is motivated by recent findings on users' predictions with high accuracy for an upcoming prediction window of hundreds of milliseconds \cite{qian_optimCell_2016}. Here, we investigate the effect of this knowledge to significantly reduce the computing latency via \emph{proactive computing}, whereby servers  proactively render the upcoming high-definition video frames which are stored at the edge server prior to  users's requests.

Ensuring a reliable link in mmwave-enabled VR environment is a daunting task since the mmWave signal experiences high level of variability and blockage. Therefore, we investigate MC as an enabler for reliable communication, in which a gaming arcade with $8$x$8$ game pods, served by multiple mmwave access points  connected to  edge servers is assumed.  We model the user association to edge servers as a dynamic matching problem to minimize service latency such that users with a link quality below a predefined threshold  are served via MC.

\begin{figure*}[t]
 \centering
\begin{minipage}{0.45\linewidth}
     \vspace{-1.35em}
	\includegraphics[width=1\columnwidth]{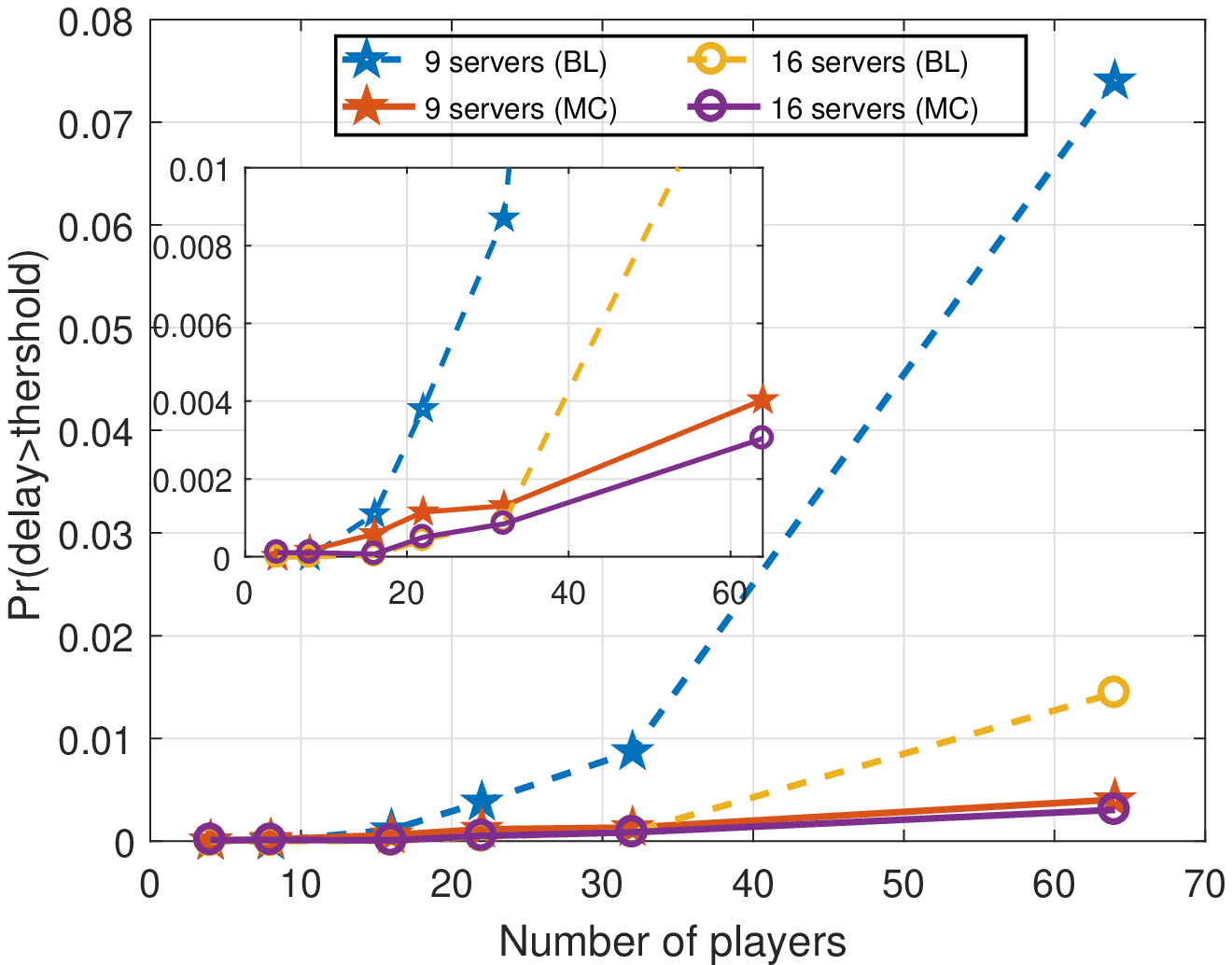}
	\caption{VR service reliability versus network density for different number of servers (BSs).}
		\label{Fig:VR-1}
\end{minipage}
\hspace{1.5em}
\begin{minipage}{0.45\linewidth}
	 \vspace{-1em}
	\includegraphics[width=1\columnwidth]{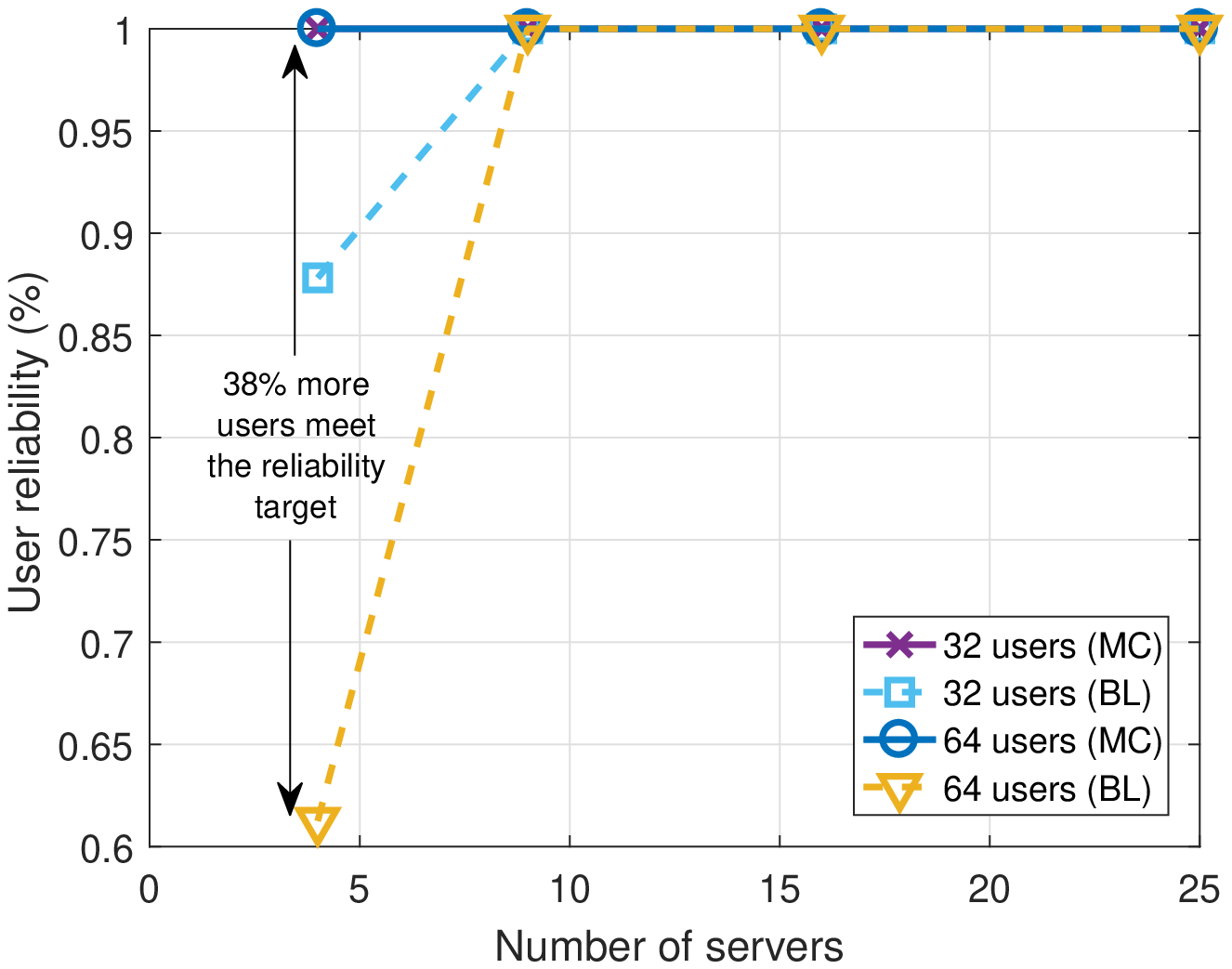}
	\caption{User reliability, expressed as the percentage of users who achieve the reliability target, versus number of servers.}
		\label{Fig:VR-2}
\end{minipage}
\end{figure*}

In Fig. \ref{Fig:VR-1}, we plot the VR service reliability of the proactive solution with MC and a baseline scheme without proactive computing nor MC. In this context, service reliability is defined as the probability of experiencing a transmission delay less than a threshold value, set here as $10$ ms.  Fig. \ref{Fig:VR-1} shows that for a given server density reliability decreases (as the rate of violating the maximum delay threshold increases) with the number of players. However, increasing the number of servers improves the network reliability, as the likelihood of finding a server with good signal quality increases. Moreover, it is shown that MC is instrumental in boosting the service reliability through overcoming mmwave signal fluctuation, and minimizing the worst service delay a user can get.  Moreover, Fig. \ref{Fig:VR-2} plots the user reliability as a function of server density. User reliability is expressed as the percentage of users who achieve the reliability target.  It can be seen that reliability by means of both proactivity and MC ensures all users are within the delay budget, even with low number of servers.

\subsection{Mobile Edge Computing}

\begin{figure*}[t]
\centering
\begin{minipage}{0.45\linewidth}
	\includegraphics[width=1\columnwidth]{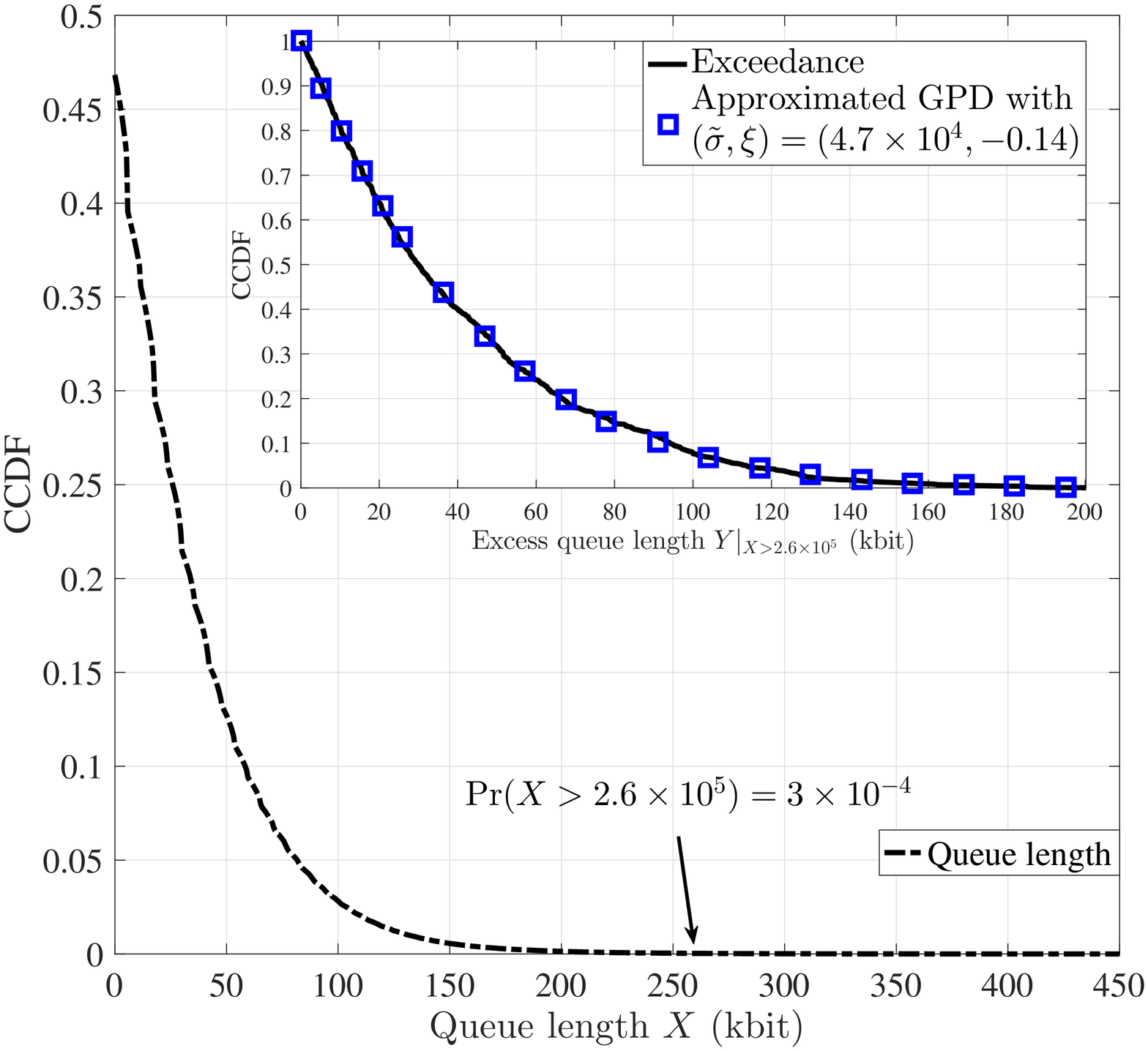}
	\caption{Tail distributions of a given UE's task queue length, queue length exceedance over threshold, and the approximated GPD of exceedances.}
		\label{Fig:EVT-1}
\end{minipage}
\hspace{1.5em}
\begin{minipage}{0.45\linewidth}
	\vspace{-1.35em}
	\includegraphics[width=1\columnwidth]{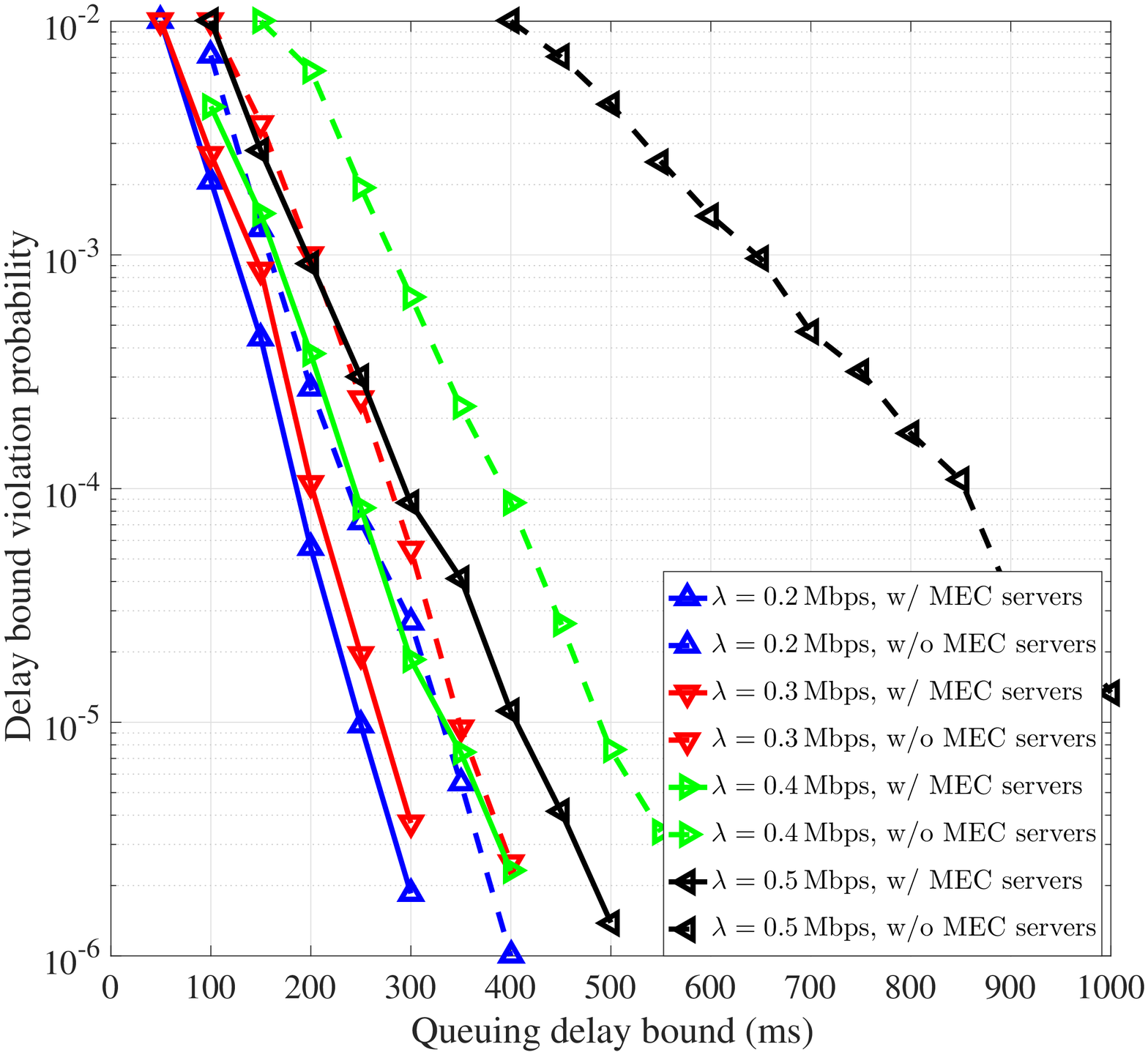}
	\caption{Delay bound violation probability versus queuing delay bound.}
		\label{Fig:EVT-2}
\end{minipage}
\end{figure*}

We consider a mobile edge computing (MEC) scenario in which  MEC servers are deployed at the network edge to provide faster computation capabilities for computing mobile devices' tasks. Although mobile devices can wirelessly offload their computation-intensive tasks to proximal MEC servers, offloading tasks incurs extra latency.
Specifically, if the number of task-offloading users is large, some offloaded tasks need to wait for the available computational resources of the servers. In this case, the waiting time for task computing at the server cannot be ignored and should be taken into account.
Since the waiting time is closely coupled with the task queue length, and extreme queue values will severely deteriorate the delay performance, we leverage extreme value theory to investigate the impact of the queue length on the performance of MEC \cite{GC17_EVT}. Here, extreme value theory allows to fully characterize the tail of the latency distribution, which is  then incorporated into the system design.
Firstly, we set a threshold $d$ for the queue length and impose a probabilistic constraint on the queue length threshold violation, i.e.,
\begin{equation}\label{Eq: ori_net_prob_cst}
\lim\limits_{T\to\infty}\frac{1}{T}\sum\limits_{t=1}^{T}{\rm Prob}(X(t)> d )\leq \epsilon,
\end{equation}
where $X(t)$ and $\epsilon\ll1$ are the queue length in time slot $t$ and tolerable violation probability.
Subsequently, we focus on the excess queue length $Y(t)=X(t)-d$ over the threshold $d$. According to Theorem \ref{Thm: GPD}, we know that the statistics of the exceedances over the threshold are characterized by the scale parameter $\tilde{\sigma}$ and shape parameter $\xi$. Thus, we formulate two constraints for the scale and shape parameters as
$\tilde{\sigma}\leq \tilde{\sigma}_{\rm th}$ and $\xi\leq \xi_{\rm th}$ which can be further cast as constraints for the mean and second moment of the excess queue value, i.e.,
\begin{align}
&\lim\limits_{T\to\infty}\frac{1}{T}\sum\limits_{t=1}^{T}\mathbb{E}\big[Y(t)|X(t)> d \big]\leq \frac{\tilde{\sigma}_{\rm th}}{1-\xi_{\rm th}},\label{Eq:GPD-1}
\\&\lim\limits_{T\to\infty}\frac{1}{T}\sum\limits_{t=1}^{T}\mathbb{E}\big[(Y(t))^2 |X(t)> d\big]\leq \frac{2(\tilde{\sigma}_{\rm th})^2}{(1-\xi_{\rm th})(1-2\xi_{\rm th})}.\label{Eq:GPD-2}
\end{align}
Utilizing Lyapunov stochastic optimization, a control algorithm is proposed for task offloading and computation resource allocation while satisfying the constraints on the queue length threshold violation \eqref{Eq: ori_net_prob_cst} and the statistics of  extreme queue length, i.e., \eqref{Eq:GPD-1} and \eqref{Eq:GPD-2}  \cite{abs-1710-00590}.

Fig.~\ref{Fig:EVT-1} shows the tail distribution, i.e., complementary cumulative distribution function (CCDF), of the queue length in which the threshold of the queue length is set as $d=2.6\times 10^5$. Given a zero-approaching  threshold violation probability, i.e., $\Pr(X>2.6\times 10^5)=3\times 10^{-4}$, and applying Theorem \ref{Thm: GPD} to the conditional excess queue value $Y|_{X>2.6\times 10^5}=X-2.6\times 10^5$, we also plot in Fig.~\ref{Fig:EVT-1} the tail distributions of the conditional excess queue value and the approximated GPD which coincide with each other.
Moreover, the shape parameter $\xi$ of the approximated GPD  allows to estimate the statistics of the maximal queue length as per Theorem \ref{Thm: GEV},  so as to proactively tackle  the occurrence of extreme events.

In order to show the impact of computation speed and task arrival rates, we vary the bound of the queuing delay and plot in Fig. 8 the delay bound violation probability as a reliability measure. Since the MEC servers provide faster computation capabilities, offloading tasks to the MEC servers further reduces the waiting time for task execution for higher computation tasks. In other words, the MEC-centric approach improves the reliability performance for delay-sensitive applications with higher computation requirements.

\subsection {Multi-connectivity for ultra-dense networks}

We investigate the fundamental problem of BS-UE association aiming at improving capacity in the context  of ultra-dense networks via multiconnectivity. By introducing a network-wide cost function $\cost(\state,\channel)$ that takes into account the tradeoff between the received signal-to-noise ratio (SNR) $\sinr$ and the network power consumption including both transmit power and the power consumption of BS and UE multiconnectivity, the network is analyzed as a function of channel statistics ($\channel$) as the number of BSs $(\BS$) and UEs ($\UE$) grows large with a given ratio $\ratio=\UE/\BS$. % under a given ratio.
Here, the optimization variable $\state$ represents the BS-UE associations.
Henceforth, the objective is to determine the optimal network-wide cost averaged over the channel statistics as follows:
\begin{equation}\label{eqn:mult_connect}
	\energy = \expect_{\channel}[\costOptimal] =
	\expect_{\channel}[\max_{\state\in\stateset}\cost(\state,\channel)] =
	\textstyle \expect_{\channel}\left[ \max\limits_{\state\in\stateset} \left(  \powerMxBS \sum\limits_b\countBS + \powerMxUE \sum\limits_{u} (\countUE-1)  - \sum\limits_{b,u}\sinr_{bu}(\state,\channel) \right) \right],
\end{equation}
where $\countBS$, $\countUE$, $\powerMxBS$, and $\powerMxUE$ are the number of UEs connected with BS $b$, number of BSs connected with UE $u$, and the power consumption for multiconnectivity at BSs and UEs, respectively.

The complexity of solving \eqref{eqn:mult_connect} grows exponentially with the number of  $\BS$ and $\UE$. Therefore, we resort  to the analysis of $\max_{\state\in\stateset}\expect_{\channel}[\cost(\state,\channel)]$ due to the reduced complexity of the solution steps instead of $\energy$.
Although the former is simpler to solve, it is oblivious to the instantaneous channel states of the network while the latter takes into account the instantaneous channel states.
Henceforth, finding an analytical expression for $\energy$ in the dense regime and obtaining important statistics of the operation of the network at this optimal point (such as the average SNR,
and  average number of connections per UE and per BS) is the prime motivation of this work.
This is achieved by using tools from statistical physics: Hamiltonian, partition sum, and replica method introduced under Section \ref{sec:mean_field_sumudu}.
Equipped with the above tools,  solving \eqref{eqn:mult_connect} is tantamount to solving a set of fixed point equations.

\begin{figure}[t]%
	\centering
	\subfloat[{Validation of the analytical expression via Monte-Carlo simulations. Optimal values of the objective function $\energy=\expect_{\channel}[\costOptimal]$ is compared for different numbers of BSs and UEs.}]
	{
		\includegraphics[width=.475\textwidth]{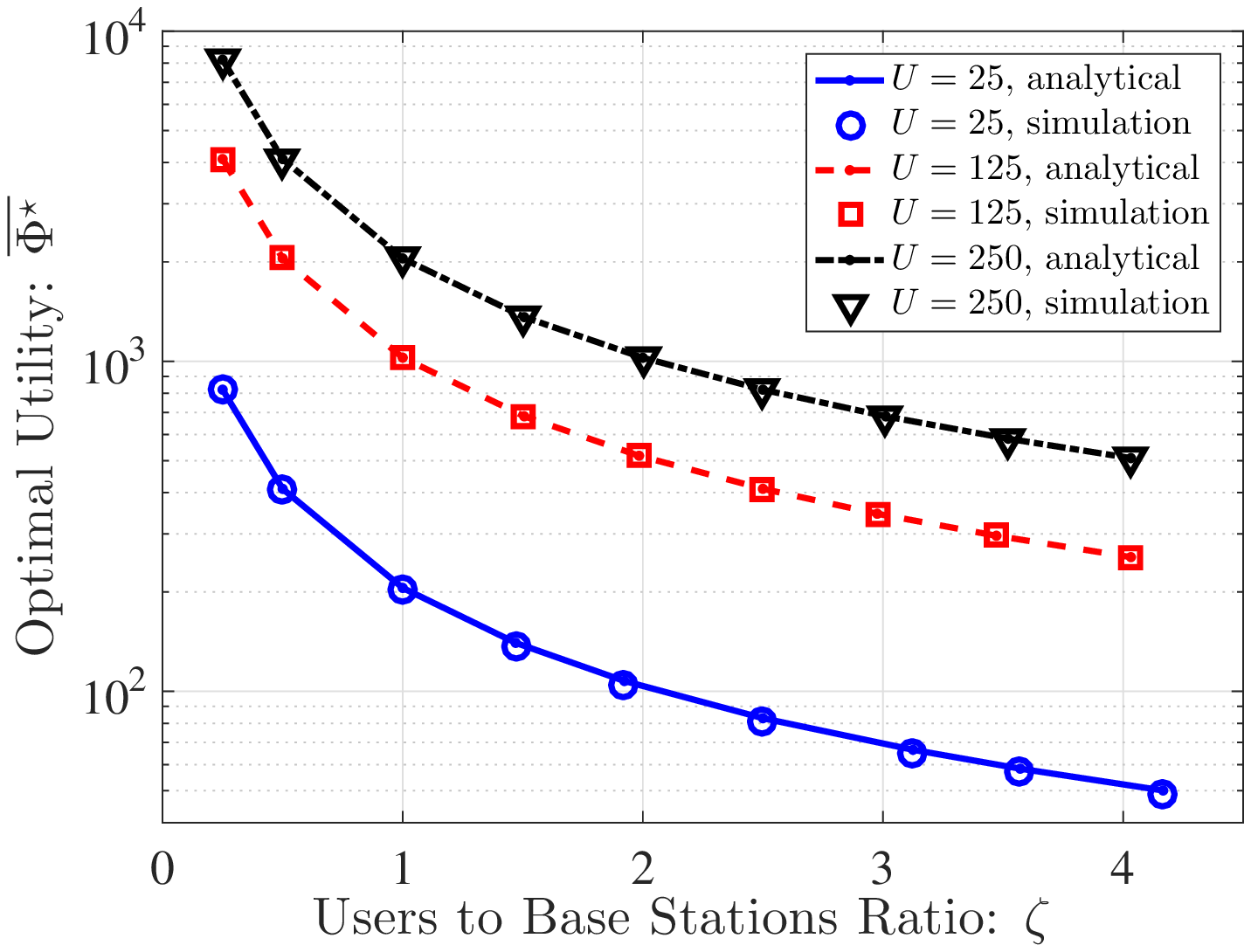}
		\label{fig_sp:validation}
	}\:
%	\subfloat[Coverage probability for different number of UEs-BSs ratios $\ratio=\UE/\BS$ with $\UE=100$. Here, the total power consumption of all BSs in the network remains fixed.]
%	{
%		\includegraphics[width=.475\textwidth]{coverage_SP}
%		\label{fig_sp:coverage}
%	}\\
	\subfloat[Reliability in terms of fraction of UEs that satisfies a given threshold for different number of UEs-BSs ratios $\ratio=\UE/\BS$ with $\UE=100$. Here, the total power consumption of all BSs  remains fixed.]
	{
		\includegraphics[width=.475\textwidth]{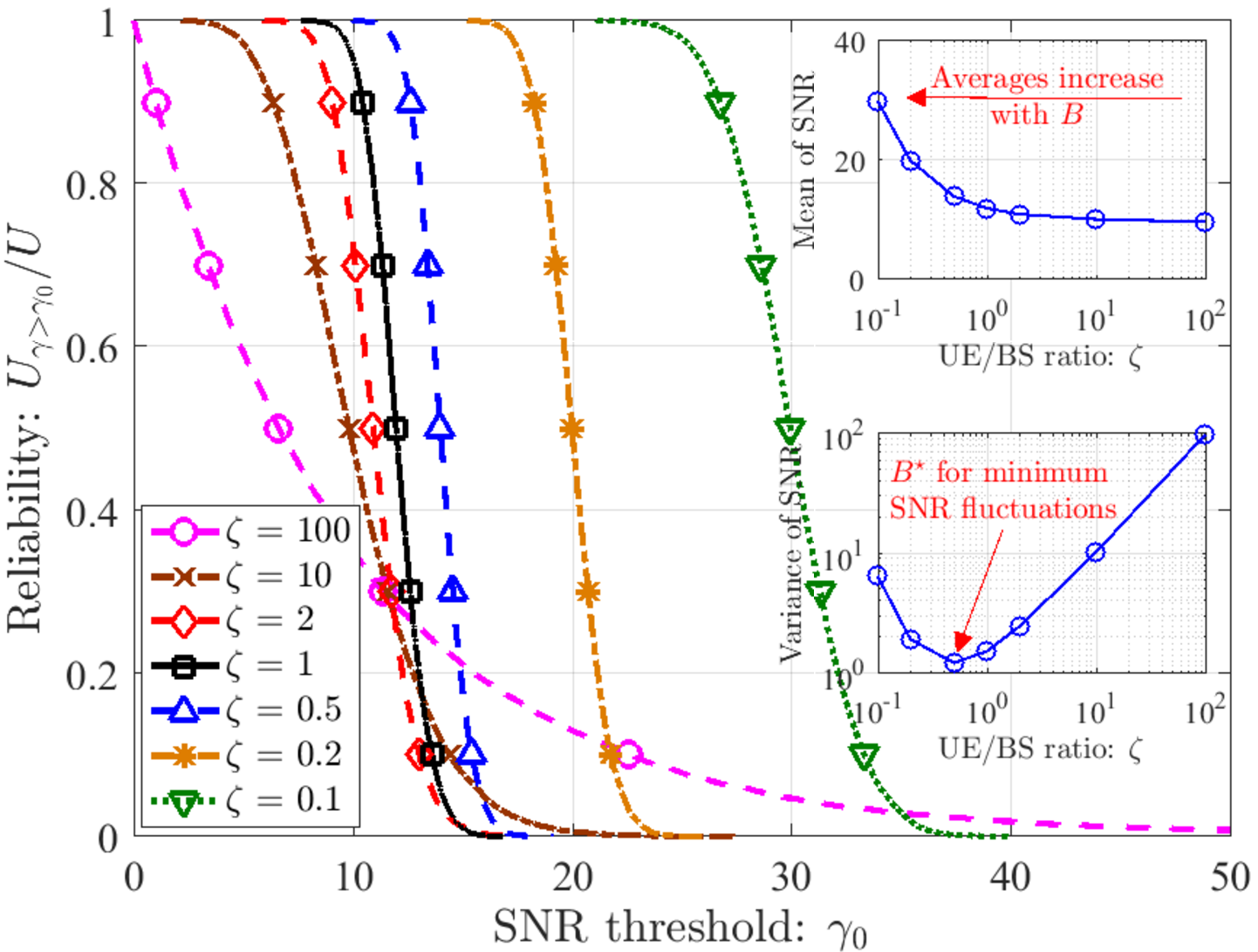}
		\label{fig_sp:reliability}
	}
	\caption{Validation of the statistical physics approach and  network performance analysis.}
	\label{fig_SP}
\end{figure}

Fig.~\ref{fig_sp:validation} validates the effectiveness of the analytical expression using extensive sets of Monte-Carlo simulations.
It can be clearly noted that the analytical results  align very well with the simulations.
This showcases the usefulness of using this methodology in gaining insights of the performance of ultra-dense networks without resorting to cumbersome and time-consuming Monte-Carlo simulators.

%The coverage probability, probability of average UE SNR is larger than a given SNR threshold $\mbox{Pr}(\sinr>\sinr_0)$, for different $\ratio$ is shown in Fig.~\ref{fig_sp:coverage}.
%For the fairness of comparison, the total power consumption of all BSs in the network has been fixed in which the comparison is among few powerful transmitters versus many low-power transmitters.
%It can be noted that the coverage increases as the number of BSs increases.
%However, the fluctuations of coverage probability is minimal only when the number of BSs are slightly higher than the number of UEs.

Fig.~\ref{fig_sp:reliability} plots the network reliability  for different $\ratio=\UE/\BS$ with $\UE=100$ users.
Here, the reliability is measured in terms of the ratio between the number of UEs that achieve an instantaneous SNR ($\sinr$) above a given threshold $\sinr_0$, $\UE_{\sinr>\sinr_0}$, to the total number of UEs ($U$).
For sake of fairness, the total power consumption of all BSs in the network is fixed and the intent is to provide insights about reliability by asking the question: ``\emph{is it better to have few powerful transmitters or  many low-power transmitters}?''.  From the figure, it can be noted that the use of many low-powered BSs provides higher reliability as compared to  a very few powerful BSs.
However, while this holds from an average perspective, fluctuations of the instantaneous users' SNRs are high for both $\ratio \ll 1$ and $\ratio \gg 1$ scenarios. This means that network topologies with $\ratio$ slightly higher than unity are suitable when the variance $\sinr$ is sought-after instead of the average $\sinr$.

\section{Conclusions}
Enabling URLLC warrants a major departure from average-based  performance towards a clean-slate design centered on tail, risk and scale.  This article has reviewed recent advances in low-latency and ultra-high reliability in which key enablers have been closely examined. Several methodologies stemming from adjacent disciplines and tailored to the unique characteristics of URLLC have been described. In addition, via selected use cases we have demonstrated how these tools provide a principled and clean-slate framework for modeling and optimizing URLLC-centric problems at the network level. This article will help foster more research in URLLC whose importance will be adamant in beyond $5$G and $6G$ networks, with the slew of unforeseen applications.

\section{Acknowledgements}
This work was supported in part by the Academy of Finland project CARMA and SMARTER,  the INFOTECH project NOOR, the Kvantum Institute strategic project SAFARI,  the 6Genesis Flagship project, and the U.S. National Science Foundation
under Grant CNS-1702808. Authors are also greatful to several people who provided their comments and suggestions, in particular Dr. Kimmo Hiltunen (Ericsson), Kari Leppanen (Huawei), Dr. Harish Viswanathan (Nokia Bell-Labs),  Dr. Chih-Ping Li (Qualcomm), and many others.
%\scriptsize
\bibliographystyle{IEEEtran}
\setlength{\itemsep}{-30mm}
\bibliography{refbib}

\begin{IEEEbiography}{Mehdi Bennis} (SM'15) is an Associate Professor at the Centre for Wireless Communications, University of Oulu, Finland and an Academy of Finland Research Fellow. His main research interests are in radio resource management, heterogeneous networks, game theory and machine learning in 5G networks and beyond. He has co-authored one book and published more than 200 research papers in international conferences, journals and book chapters. He also has organized a dozen of IEEE workshops in various topics (small cells, UAVs and VR). He has been the recipient of several awards including the 2015 Fred W. Ellersick Prize from the IEEE Communications Society, the 2016 Best Tutorial Prize from the IEEE Communications Society, the 2017 EURASIP Best paper Award for the Journal of Wireless Communications and Networks and the all-University of Oulu award for research. Dr Bennis is currently an editor of IEEE TCOM.
 \end{IEEEbiography}

\begin{IEEEbiography}{Merouane Debbah} (F'15) entered the Ecole Normale Supérieure Paris-Saclay (France) in 1996 where he received his M.Sc and Ph.D. degrees respectively. He worked for Motorola Labs (Saclay, France) from 1999-2002 and the Vienna Research Center for Telecommunications (Vienna, Austria) until 2003. From 2003 to 2007, he joined the Mobile Communications department of the Institut Eurecom (Sophia Antipolis, France) as an Assistant Professor. Since 2007, he is a Full Professor at CentraleSupelec (Gif-sur-Yvette, France). From 2007 to 2014, he was the director of the Alcatel-Lucent Chair on Flexible Radio. Since 2014, he is Vice-President of the Huawei France R$\&$D center and director of the Mathematical and Algorithmic Sciences Lab. His research interests lie in fundamental mathematics, algorithms, statistics, information and communication sciences research. He is an Associate Editor in Chief of the journal Random Matrix: Theory and Applications and was an associate and senior area editor for IEEE Transactions on Signal Processing respectively in 2011-2013 and 2013-2014. Mérouane Debbah was the recipient of the ERC grant MORE (Advanced Mathematical Tools for Complex Network Engineering) from 2012 to 2017. He is a IEEE Fellow, a WWRF Fellow and a member of the academic senate of Paris-Saclay. He has managed 8 EU projects and more than 24 national and international projects. He received 19 best paper awards for his research publications. \end{IEEEbiography}

\begin{IEEEbiography}{H.  Vincent  Poor} (S'72, M'77, SM'82, F'87) is  the  Michael  Henry  Strater  University  Professor  of Electrical Engineering at Princeton University.  From 1977 until he joined the Princeton faculty in  1990,  he  was  a  faculty  member  at  the  University  of  Illinois  at  Urbana-Champaign.  During 2006-16, he served as Dean of Princeton’s School of Engineering and Applied Science. He has also held visiting appointments at several other universities, including most recently at Stanford and Cambridge. His research interests are primarily in the areas of information theory, statistical signal  processing  and  stochastic  analysis,  with  applications  in  wireless  networks and  related fields.    Among  his  publications  in  these  areas  is  the  recent  book \emph{Information Theoretic Security and Privacy of Information Systems} (Cambridge University Press, 2017). Dr.  Poor  is  a  member  of  the  National  Academy  of  Engineering  and  the  National Academy  of Sciences,  and  is  a  foreign  member  of  the  Chinese Academy of Science, the Royal Society, and other national and international academies.  He  has  served  as  the  President  of  the IEEE   Information   Theory   Society,   and   as   Editor-in-Chief   of   the \emph{IEEE   Transactions   on Information Theory}. In 2002 he received a Guggenheim Fellowship, and in 2005 he received the IEEE  Education  Medal.  Recent  recognition  of  his  work  includes the 2014  URSI  Booker  Gold Medal, the 2016 John Fritz Medal, the 2017 IEEE Alexander Graham Bell Medal, and honorary degrees and professorships from a number of universities worldwide.
\end{IEEEbiography}

\end{document}